\documentclass{aa}
\usepackage[english]{babel}
\usepackage{graphics}

\newcommand{\beq}{\begin{equation}}
\newcommand{\eeq}{\end{equation}}
\newcommand{\bea}{\begin{eqnarray}}
\newcommand{\eea}{\end{eqnarray}}
\newcommand{\tensor}[1]{\buildrel\leftrightarrow\over #1}

\begin{document}
\thesaurus{09(02.13.2; 08.16.5; 08.13.2; 09.10.1)}
\title{Collimated outflows of rapidly rotating young stellar objects}
\subtitle{Wind equation, GSS equation and collimation}
\author{Elena Breitmoser \and Max Camenzind}
\institute{Landessternwarte K\"onigstuhl,
D--69117 Heidelberg, Germany}
\offprints{Elena Breitmoser}
\mail{ebreitmo@lsw.uni-heidelberg.de}
\date{Received \today; accepted \today}
\titlerunning{Collimated outflows of young stellar objects}
\authorrunning{E.~Breitmoser \& M.~Camenzind}
\maketitle

\begin{abstract}
The magnetic field structure and the collimated outflow of rapidly rotating Young Stellar Objects (YSOs) are calculated from the stellar source to the asymptotic region. The calculations are based on ideal MHD and the further simplifying assumptions of stationarity and axisymmetry. The star--disk--jet system can be completely defined by the Grad-Shafranov (GSS) equation, describing the structure of the magnetospheres, and the wind equation, which are given by equilibrium perpendicular and parallel to the field lines. Both equations must be solved simultaneously to obtain a self-consistent solution. General solutions of the Grad-Shafranov equation are not yet available. Here we discuss an analytical model for the magnetic flux surfaces which is a solution for small and large radii. This model assumes a stellar dipole field and a gap between the star and a disk at a distance of a few stellar radii. Due to the features of the disk no field can penetrate the disk and the resulting opening of the field lines close to the polar cap is obtained as a computational result. In addition, our model guarantees the collimation of the outflow into a cylindrical shape at asymptotic jet radii of several thousand stellar radii. This model for collimated outflows reproduces all essential properties of magnetospheres for rapidly
rotating stars.
The result is used as input to the wind equation. This problem is completely integrable, determined by five constants of motion: the total energy E, the total angular momentum L, the total mass flux $\eta$ along a flux surface, and the total entropy S in a flux surface, together with the rotation $\Omega^F$ of the field lines. For adiabatic plasma flows, this problem is algebraic and can easily be solved. Pressure is neglected in our computation.

The theory of axisymmetric magnetospheres around rapidly rotating stellar
sources is outlined including electric fields due to the rapid rotation.
Gravity of the central object is consistently built into this theory.
Due to the injection of plasma either from the stellar surface or by
interaction with a surrounding disk, these magnetospheres are neither vacuum 
solutions, nor force--free. A consistent wind theory is developed which
contains the Newtonian theory as a classical limit. 

Current--carrying plasma flows will lead to a collimation of the magnetospheric
structure into a cylindrical shape. Particular solutions are discussed for
the asymptotic collimation. We show that the asymptotic structure is essentially determined by electric forces in the pinch equation, and not by centrifugal
and pressure forces.
\end{abstract}

\section{Introduction}

The phenomenon  of highly collimated high-velocity bipolar outflows occurs frequently in Astrophysics for a large range of objects such as protostars and Active Galactic Nuclei. The 
formation of jets seems to be closely related to the presence of magnetic fields and the 
existence of a gaseous disk around a rapidly rotating central object (Camenzind 1990). It is widely believed that due to these rotating magnetospheres, the mass flow can be accelerated and collimated on scales of a few tens of AU as shown by HST observations of Young Stellar Objects (\cite{Ray}). Nevertheless, their origin is still under discussion. Protostellar jets are most probably disk winds or stellar winds. The structure of these magnetospheres follows from solutions 
of the Grad-Shafranov equation for force--balance perpendicular 
to the field lines. 
 Magnetospheres generated by isolated stars are typically
three dimensional objects (e.g., the Earth's magnetosphere, pulsar magnetospheres or stellar magnetospheres) and as such are very complicated to model consistently. There are a few
types of systems where the magnetosphere is constrained to be axisymmetric.
Accretion disks will produce highly axisymmetric structures and stars
in close interaction with accretion disks are also sources of axisymmetric
magnetospheres. Magnetized rotators are natural sources of energetic plasma flows.
We investigate the dynamics of the wind in these 
collimated magnetospheres for axisymmetric and stationary flows. The shape of 
the magnetic flux function determines whether the wind is accelerated or decelerated. 

In this paper we develop a general theory for steady axisymmetric ideal
plasma flows driven by rapidly rotating magnetospheres of stellar sources
including contributions from electric fields. This theory also includes
gravity, such that outflows driven by neutron stars and black holes are covered.
With the elements developed in this paper one can calculate plasma flows
driven by young stellar sources, where the central object is rotating near its
breakup velocity, as well as winds driven away by rapidly rotating neutron
stars. This is an extension of the work by Mobarry \& Lovelace (1986)
including new insights into this problem from the last ten years.

An overview of the attempts to approach the problem of solving the Newtonian Grad-Shafranov equation and the wind equation can be found in \cite{lery98}. \cite{lery98}  also present a model for stationary, axisymmetric MHD winds without the need for self-similar assumptions as is done here.
 Whereas our model gives the structure of the magnetospheres for all regions, they find a solution of the Grad-Shafranov equation for the Alfv\'en surface from the  Alfv\'en regularity condition and at the base of the flow. Their magnetic field structure for small radii is conical in shape, which does not take a stellar dipole field or the influence of a disk into consideration. We find that the shape of the flux surfaces close to the stellar source is important for the acceleration of the plasma. The Alfv\'en point and the fast magnetosonic point are rather close to each other. The fast magnetosonic surface lies at radii smaller than the inner disk radius.
The collimation of the magnetospheres into a cylindrical outflow is obtained by imposing an external pressure in \cite{lery99}. That is not necessary for our model.
When plasma flows reach radii near the light cylinder, electric forces 
become important for the modelling of the underlying magnetosphere for fast rotators (Camenzind 1986a; Fendt et al. 1995; Fendt \& Camenzind 1996). This does not mean that plasma flows are accelerated
to relativistic energies, only that electric fields
contribute to the forces as much as magnetic fields do. The electric field
can only be neglected, at radii much smaller than the
light cylinder radius (at least only a few percent of the light cylinder).
The poloidal current follows from the MHD wind theory.
For \cite{lery99} the light cylinder is of no importance.
In \cite{lery98} the wind equation is solved from the source to the fast magnetosonic point, whereas our wind solution is global and goes up to the asymptotic region of the jet. In contrary to our cold wind solution they use a polytropic equation of state. For a cold wind our fast magnetosonic surface does not go to infinity, because we impose a dipole--like structure on the flux surfaces instead of a monopole--like one. In addition, their wind solution represents the Newtonian limit of our derivation.
Newtonian MHD is perfect for modelling the magnetic structure of the solar
wind (Mac Gregor 1996). But winds ejected from protostars rapidly rotating with periods of a few days
could pass the light cylinder given as (Camenzind 1997)
\beq
  R_L = {c\over\Omega^F} = 27\,\rm AU\,{P_*\over{days}} \,.
\eeq
Such a radius is interestingly close to the observed collimation radii.
While the light cylinder is not essential for the plasma kinematics, it does influence the collimation processes at large radii by means of electric fields due to field line rotation.

The paper is structured as follows.
In Sect. \ref{chap:2} a consi\-stent wind theory for rapidly rotating objects is presented. Axisymmetric fields and stationary flows are investigated. The Newtonian case follows quite naturally as a limit of the special relativistic MHD.
The polytropic equation of state is included. A general expression for the wind equation is given and the special solution for a cold wind is discussed. In Sect. \ref{chap:3} the Grad-Shafranov equation is studied. Sect. \ref{chap:4} gives a discussion of the collimation processes for spherical outflows. The importance of the electric fields close to the light cylinder is shown.
In Sect. \ref{chap:5} a model is presented to reproduce the characteristics of collimated outflows near to and far from the star. A calculation of the magnetic flux surfaces and the resulting wind is given in Sect. \ref{chap:6} for a typical young, low-mass star. Also the influence of various parameters of the model is studied. We draw our conclusions in the last section.

\section{A consistent wind theory for rapidly rotating stars}\label{chap:2}
A modern treatment of magnetohydrodynamics including the effects of gravity
must be based on a general relativistic approach in the 3+1 split 
of space--time (see e.g. \cite{thorne}; Beskin \& Pariev 1993; 
Camenzind 1996). When the rotation of the central object is not important, 
the line element is given by the Schwarzschild decomposition
\beq
  ds^2 = - \alpha^2\,c^2\,dt^2 + {1\over\alpha^2}\,dr^2 + r^2\,d\theta^2
    + r^2\sin^2\theta\,d\phi^2\,.
\eeq
Here $\alpha = \sqrt{1 - 2GM_*/c^2r}$ is the redshift factor which accounts for
the gravitational redshift between a local observer and infinity. It describes the effects of gravity due to the presence of the central star of mass $M_*$. In the Newtonian limit we have
\beq
  \alpha \simeq 1 + \Phi/c^2\,,
\eeq
where $\Phi$ is the Newtonian potential. In the following we also use the
cylindrical radius $R = r \sin\theta$. The time $t$ is the global time.

The spin $J_*$ of the central object is also a source of gravity in general
relativity (the so--called gravitomagnetic effect). This spin would produce an
off--diagonal element $g_{t\phi}$ in the above line element, 
$g_{t\phi} \propto J_*/r^3$, which can be important for neutron stars
rotating with millisecond periods (Camenzind 1986b). This effect is
crucial for the treatment of plasma flows on rapidly rotating black holes
(Camenzind 1996). Since we are interested here for applications to stellar
sources, we will neglect these effects.

Local fiducial observers (FIDOs) are represented by orthonormal tetrads
\bea
  O &=& {1\over{\alpha c}}\,\partial_t \quad,\quad 
    \vec{e}_r = \alpha\,\partial_r\, , \\
  \vec{e}_\theta &=& {1\over r}\,\partial_\theta \quad,\quad
    \vec{e}_\phi = {1\over R}\,\partial_\phi \,.
\eea    
    
\subsection{3+1 split of electrodynamics}
On a curved spacetime background electromagnetic fields must be given in terms of
the Faraday tensor $F^{\mu\nu}$, and Maxwell's equations are given in
tensorial form. With respect to the particular observer field defined
above, we can however split the Faraday tensor as in special relativity
into an electric and magnetic part
\beq
  F^{\mu\nu} = O^\mu E^\nu - E^\mu O^\nu + \epsilon^{\mu\nu\rho\sigma}\,
    O_\rho B_\sigma \,.
\eeq
Similarly, we get a charge density
\beq
  \rho_e = - \vec{O}\cdot\vec{J}
\eeq
and a current density\footnote{We use the notation of Thorne et al. 1986 for the four--vectors.} $\vec{\cal J}$ given by
\beq
  \vec{\cal J} = \rho_e O + J\,.
\eeq

In terms of the electric fields $\vec{E}$ and magnetic fields $\vec{B}$,
Maxwell's equations assume the familiar form (\cite{thorne})
\bea
  \nabla\cdot\vec{E} &=& 4\pi\rho_e \, ,\\
  \nabla\cdot\vec{B} &=& 0\, , \\
  \nabla\times (\alpha\vec{E}) &=& - {{\partial\vec{B}}\over{c\partial t}}\, ,\\
  \nabla\times (\alpha\vec{B}) &=& + {{\partial\vec{E}}\over{c\partial t}}
   + {{4\pi}\over c}\,\alpha\vec{\cal J} \,.
\eea
Charge conservation is given as
\beq
  {{\partial\rho_e}\over{\partial t}} + \nabla\cdot(\alpha\vec{\cal J}) = 0 \,.
\eeq

In this treatment, the redshift factor $\alpha$ only appears as a form factor, but there are no additional couplings between electromagnetic fields and
gravitational forces. This is in contrast to the treatment on rapidly
rotating backgrounds, where the gravitomagnetic force couples directly
into Maxwell's equations (Camenzind 1996).

\subsection{Axisymmetric fields}
As in the Newtonian theory of plasma confinement, the above equations
can be considerably simplified for axisymmetric structures. Here, we can define
the poloidal flux function $\Psi(r,\theta)$
\beq
  \Psi(r,\theta) = {1\over{2\pi}}\,\int_A \vec{B}_p\cdot d\vec{A}\,,
\eeq
and the axial current flowing downward along the central axis
\beq
  I(r,\theta) = -\int_A \alpha\vec{j}_p\cdot d\vec{A} = - {c\over 2}\,
    \alpha RB_T\,.
\eeq
$A$ denotes an upward directed surface around the central axis. The level
surfaces of $\Psi$ are the magnetic surfaces which are generated by rotating
field lines. The level surfaces of $I$ are the surfaces where the poloidal
currents flow.

Due to stationarity and axisymmetry, the electromagnetic fields assume a simple
form 
\bea
  \vec{B} &=& {1\over R}\,\nabla\Psi\times\vec{e}_\phi - 
    {{2I}\over{R\alpha c}}\,\vec{e}_\phi, \\
  \vec{E}_p &=& - {\Omega^F(\Psi)\over{\alpha c}}\,\nabla\Psi 
    \quad ,\quad \vec{E}_T = 0\,.
\eea
Infinite conductivity requires $\vec{B}\cdot\vec{E} = 0$, and thus the
integration constant $\Omega^F(\Psi)$ represents the angular velocity of
field lines. This is not the angular velocity of a plasma particle.    
             
\subsection{Plasma motion}
We treat plasma motion only in the one--fluid approach (see \cite{khanna98}).
For a one--component
non--viscous plasma the energy--momentum tensor has the form
\beq
  T = (\rho + P)\,u\otimes u + Pg + T_{\rm em}\, ,
\eeq
$\rho$ is the total energy of the plasma including internal energy, $P$
the proper pressure, $u$ the four--velocity of the plasma and $g$ the metric tensor.
We decompose the 
energy--momentum tensor $T$ into horizontal and vertical components with respect to fiducial observers
\beq
  T = \epsilon\,O\otimes O + O\otimes\vec{S} + \tensor{t}\,.
\eeq
$\epsilon$ is the energy--density with respect to local observers, 
$\vec{S}$ the momentum flux and $\tensor{t}$ the stress tensor. 

With the decomposition of $u = \gamma(O + \vec{v})$ we obtain the following 
components with respect to FIDOs
\bea
  \epsilon &=& \gamma^2\,(\rho + P\vec{v}^2) + \epsilon_{\rm em}\, ,\\
  \vec{S}  &=& (\rho + P)\,\gamma^2\vec{v} + \vec{S}_{\rm em} =
    \rho_0 \mu \gamma^2\,\vec{v} + \vec{S}_{\rm em}\, , \\
  \tensor{t}  &=& (\rho + P)\,\gamma^2\,\vec{v}\otimes\vec{v} + P\tensor{g} + 
    \tensor{t}_{\rm em} \nonumber\\
   &=& \vec{S}\otimes\vec{v} + P\tensor{g} + 
    \tensor{t}_{\rm em} \,.
\eea
Here, $\rho_0$ is the rest mass density, $\mu = (\rho + P)/n$ the relativistic
specific enthalpy and $\gamma$ the Lorentz factor measured by fiducial observers. $\vec{S}_{em}$ represents the Poynting flux measured by 
FIDOs and $\tensor{t}_{\rm em}$ the Maxwell stresses
\bea
  \epsilon_{\rm em} &=& {1\over{8\pi}}\,\Bigl(\vec{E}^2 + \vec{B}^2\Bigr)\, ,\\
  \vec{S}_{\rm em}  &=& {1\over{4\pi}}\,\vec{E}\times\vec{B}\, , \\
  \tensor{t}_{\rm em}  &=& {1\over{4\pi}}\,\Bigl( -\vec{E}\otimes\vec{E} -
    \vec{B}\otimes\vec{B} + {1\over 2}\tensor{g}(\vec{E}^2 + \vec{B}^2) \Bigr)\,.\label{equ:25}
\eea 
Using the 3+1 split of the affine connection,    
one can now derive the 3+1--split of the hydrodynamic equations 
$\nabla\cdot T = 0$. 
Energy conservation, given by $O\cdot(\nabla\cdot T) = 0$, can be written as (\cite{thorne})
\beq
  {{d\epsilon}\over{d\tau}} = 
  {1\over\alpha}\,\partial_t\,\epsilon =
    - {1\over\alpha^2}\,\nabla\cdot(\alpha^2\vec{S}) \,.\label{energy}
\eeq
The right--hand side in the energy conservation is the familiar divergence of the energy flux, with one factor of $\alpha$ inside the divergence to account for
the gravitational redshift of the energy and the other to convert the differential proper time element in the definition of the flux into to a differential universal time element. 

Similarly, Euler's equations, given by $h\cdot(\nabla\cdot T) = 0$, where $h$ is the projector into the plasma rest frame, assume the form
\beq
  {{d\vec{S}}\over{d\tau}} = 
  {1\over\alpha}\,\partial_t \vec{S} =
   - \epsilon\nabla(\ln\alpha) - {1\over\alpha}\,\nabla\cdot(\alpha\tensor{t})
    \,.\label{momentum}
\eeq
The quantity $-\nabla\ln\alpha$ represents the local gravitational force measured by observers.    
          
\subsection{Stationary flows}
The plasma flow is given by the continuity equation
\begin{equation}
  \nabla\cdot(\alpha\gamma n\vec{v}) = 0 \,.
\end{equation}
Similarly to the Newtonian theory, one also obtains for the poloidal velocity
\begin{equation}
  \vec{u}_p = \gamma\vec{v}_p = {\eta\over{\alpha n}}\,\vec{B}_p\,,
\end{equation}
and, from the frozen--in condition, that the plasma is guided by the structure of the axisymmetric flux surfaces 
$\Psi = const$
\begin{equation}
 \fbox{$ \displaystyle
  \vec{u} = {\eta(\Psi)\over{\alpha n}}\,\vec{B} + 
    {{R\gamma\Omega^F}\over\alpha}\,\vec{e}_\phi \,.$}\label{equ:30}
\end{equation}
As a consequence of Faraday's induction law, the rotation of the 
field lines remains constant at $\Omega^F(\Psi)$. The above relation tells us
that the particle flux in a flux surface remains constant.

For stationary and axisymmetric flows the energy equation (\ref{energy})
is now simply given as
\begin{equation}
  \nabla\cdot(\alpha^2\vec{S})  = 0 \,,
\end{equation}
and similarly for the momentum equation (\ref{momentum})
\begin{equation}
  {1\over\alpha}\nabla\cdot(\alpha\tensor{t}) + \epsilon\nabla\ln\alpha
    = 0\,.
\end{equation}
By means of the MHD relations the energy flux can be written as
\begin{equation}
  \vec{S} = \mu\gamma\,{\eta\over\alpha}\,\vec{B}_p + {{I \Omega^F}\over
    {2\pi\alpha^2}}\,\vec{B}_p \,.
\end{equation}
In the stationary case, this provides the following expression for the
energy equation    
\begin{equation}
  \nabla\cdot\left( \mu\gamma\alpha\eta\,\vec{B}_p +
    {{I\Omega^F}\over{2\pi}}\,\vec{B}_p \right) = 0\,.
\end{equation} 
Using flux conservation $\nabla\cdot\vec{B}_p = 0$ and $(\vec{B}_p\cdot\nabla)
\eta = 0$ we find conservation for the total energy per particle
\begin{equation}
 \fbox{$ \displaystyle
  E(\Psi) = \mu\alpha\gamma + {{\Omega^FI}\over{2\pi\eta}}\,. $}
\end{equation}

A similar conservation law follows from the angular momentum equation 
\begin{equation}
  \vec{e}_\phi\cdot\nabla\cdot(\alpha \tensor{t}) = 0\,.
\end{equation}
Due to axisymmetry, all other terms drop out. Since
\bea
 \vec{e}_\phi\cdot \tensor{t} &=& t_{p\phi} = \mu \vec{u}_pj - 
   {1\over{4\pi}}\,\vec{B}_pB_\phi \nonumber\\
   &=& {{\mu\eta}\over\alpha}
    \,\vec{B}_pj + {I\over{2\pi}}\,\vec{B}_p
\eea
we get
\begin{equation}
  \nabla\cdot\left( \mu\eta j\vec{B}_p + {I\over{2\pi}}\vec{B}_p \right)
  = 0\,,
\end{equation}
or the total angular momentum conservation
\begin{equation}
 \fbox{$ \displaystyle
  L(\Psi) = \mu j + {I\over{2\pi\eta}} \,,$}
\end{equation}
with the specific angular momentum $j=R^2 \Omega/\alpha^2$.
In the total energy, the kinetic part is redshifted.
Together with the above relation (\ref{equ:30}) for the toroidal motion,
\begin{equation}
  u_\phi = {\eta\over{\alpha n}}\,B_\phi + {{R\gamma\Omega^F}\over\alpha} \,,
\end{equation}
this provides three equations for the unknowns $\alpha\gamma$, $j$ (or $u_\phi$) and $I$.            

In addition to these relations, an adequate equation of state is needed.
For $n = n(P,s)$ and $T = T(P,s)$ the first law of thermodynamics implies
\begin{equation}
  d\mu = {1\over n}\,dP + T\,ds\,.
\end{equation}
In particular, for $P = K_0(s)\,n^\Gamma$ with a polytropic index $\Gamma$,
one finds for the specific enthalpy
\begin{equation}
  \mu = mc^2 + {\Gamma\over{\Gamma - 1}}\,K_0(s)\,n^{\Gamma - 1} \,.
\end{equation}
With this expression, we can derive the Newtonian limit for the total energy,
$\alpha = 1 + \Phi/c^2$, $\gamma = 1 + \vec{v}^2/2\,c^2$
\begin{equation}
  E = mc^2 + {1\over 2}\,\vec{v}^2 + {\Gamma\over{\Gamma -1}}\,{P\over\rho_0}
  + \Phi + {{I\Omega^F}\over{2\pi\eta}}\,,\label{equ:46}
\end{equation}  
which is the standard energy conservation used in Newtonian MHD.

\subsection{Lorentz factor, angular momentum and poloidal current}
For given integrals of motion $E(\Psi)$, $L(\Psi)$, $\Omega^F(\Psi)$, 
$\eta(\Psi)$ and $K(\Psi)$ the
conservation laws for stationary and axisymmetric plasma flows can be reduced
to give relations for the Lorentz factor, the angular momentum and the
poloidal current flux function $I(R,\Psi)$
\begin{eqnarray}
  \alpha\gamma & = & \alpha\gamma(\Psi, R) = {E\over\mu}\,
    {{\alpha^2(1 - \epsilon) - M^2}\over
     {\alpha^2 - (R/R_L)^2 - M^2 }} \,,\label{equ:47}\\
  j & = & j(\Psi, R) = {E\over{\mu}}\,R_L\, c
    {{(1 - \epsilon)(R/R_L)^2 - M^2\epsilon}\over           
     {\alpha^2 - (R/R_L)^2 - M^2 }} \,,\label{equ:48}\\
  I & = & I(\Psi, R) = {{2\pi\eta E}\over{\Omega^F}}
     \,{{\alpha^2\epsilon - (R/R_L)^2}\over{\alpha^2 - (R/R_L)^2 - M^2 }} \,\label{equ:49},
\end{eqnarray}     
where the quantity $M$, defined as
\begin{equation}
 \fbox{$ \displaystyle
  M^2 = {{4\pi\mu\eta^2}\over n} = {{\alpha^2\,u_p^2}\over u_A^2} = {{\alpha\,u_p\,x^2}\over {c\,\sigma\,\Phi}}\,,$}\label{equ:50} 
\end{equation}
represents the (poloidal) Alfv\'en Mach number. We normalize radii in units of the light cylinder $R_L$, $x=R/R_L$. The redshift factor $\alpha$ 
corrects
for the singular behaviour of the poloidal velocity at the horizon of the
black hole. \footnote{These relations have been derived for the first time in 
Camenzind (1986b) without using 3+1 split.}
The Alfv\'en Mach number $M$ is a well--behaved quantity, e.g. at the
horizon. The quantity $\epsilon$ defined by
\begin{equation}
  \epsilon(\Psi) \equiv {{\Omega^FL}\over E}
\end{equation}
is a measure for the amount of energy carried by the electromagnetic fields.
In special relativity, $\epsilon \le 1$, and $\epsilon \equiv R_A^2/R_L^2$ is a
direct measure of the position of the Alfv\'en surface in terms of the
light cylinder radius $R_L$. For relativistic flows, $R_A \rightarrow R_L$,
and therefore $\epsilon \rightarrow 1$.
For the solar wind $\epsilon_\odot \simeq 10^{-8}$ and for winds ejected
by rapidly rotating young stellar objects we find $\epsilon_* \simeq 10^{-6}$.
In the case of the Crab pulsar $\epsilon_p \simeq 0.999$ and the wind dynamics is essentially relativistic.

This quantity $\epsilon$ determines the position of the 
{\it Alfv\'en surface}
(positions, where the denominators in the above equations vanish)
\begin{equation}
  \alpha^2(R_A) - (R_A/R_L)^2 = M_A^2 \,.
\end{equation}
This equation has in general, for given rotation $\Omega^F$, only one solution,
the outer Alfv\'en surface corresponding to the special relativistic one.
The inner Alfv\'en surface only exists in black hole magnetospheres. The light
cylinder surfaces are special solutions of this equation for $M_A^2 = 0$.

The regularity condition at the Alfv\'en point also determines the total 
angular momentum $L$ as a function of the position of the Alfv\'en point
\begin{equation}
  M_A^2\,{L\over E} = R_A^2\Omega^F \,(1 - \epsilon)\,,
\end{equation}
which becomes in the Newtonian case simply $L = R^2_A\Omega^F$. 

\subsection{The Newtonian limit}
There is much confusion in the literature as to the Newtonian
limit of the special relativistic MHD.
The two equations (\ref{equ:48}) and (\ref{equ:49}) are well--known in Newtonian MHD ($\alpha = 1$,
$\epsilon \ll 1$ and $R_L\rightarrow\infty$), 
where they form the basis for a 
treatment of axisymmetric MHD winds in stellar systems (\cite{heyvaerts96})
\begin{eqnarray}
  j & = & R^2\Omega = {{R^2\Omega^F - M^2L}\over{1 - M^2}} \,,\label{equ:51}\\
  RB_\phi & = & - 4\pi\eta\,{{L - R^2\Omega^F}\over{1 - M^2}} \,.\label{equ:52}
\end{eqnarray}  
In Newtonian MHD, the Mach number is simply $M^2 = v_p^2/v_A^2$. The first
equation indicates that plasma is corotating with the field lines for low
Mach numbers, $M^2 \ll 1$, and that the specific angular momentum $j$ is equal 
to the total angular momentum $L$ for high Mach numbers, $M^2 \gg 1$.
These equations tell us that Newtonian MHD is only valid inside the
light cylinder $R_L = c/\Omega^F$. Since for stellar objects, $\epsilon \ll 1$, a comparison between Eqs. (\ref{equ:48}--\ref{equ:49}) and (\ref{equ:51}--\ref{equ:52}) tells us that the Newtonian limit is correct for the wind equation, as long as $M^2 \gg x^2$ for all relevant radii. This approximation is therefore not
justified e.g. for rapidly rotating magnetic surfaces generated by accretion disks
around black holes. Even in the case of magnetic surfaces
generated by rapidly rotating young stars slight modifications occur in the wind equation.

In the Newtonian limit, the energy conservation at the injection point can
be written as
\beq
  {E\over{mc^2}} - 1 = {\Phi_*\over c^2} + {1\over 2}\,{\vec{v}_*^2\over c^2} + 
  {\Gamma\over{\Gamma - 1}}\,{P_*\over{mc^2n_*}} + 
  {{I_*\Omega^F}\over{2\pi m\eta c^3}}.
\eeq
Using the solution for the current at the injection point, $x_*^2 \ll \epsilon$,
$M_*^2 \ll 1$ and $p_* = kT_*/mc^2$,
\beq
  I_* \simeq {{2\pi\mu\eta}\over\Omega^F}\, {\epsilon\over{\epsilon - 1}}\,,
\eeq
this shows the various contributions to the total energy
\beq
  {E\over{mc^2}} - 1 = {\Phi_*\over c^2} + {1\over 2}\,{\vec{v}_*^2\over c^2} + 
  {\Gamma\over{\Gamma - 1}}\,p_* + {\epsilon\over{\epsilon - 1}}\,.\label{equ:ener}
\eeq
The parameter $\epsilon$ is therefore a real measure of the Poynting flux
at the injection point in units of rest mass energy flux. For the solar wind
one finds $\epsilon_\odot = 10^{-8}$ and $p_* = 10^{-6}$, and therefore the
solar wind is thermally driven and not Poynting flux driven (see also
MacGregor 1996). For rapidly rotating protostars, the Poynting flux
increases, $\epsilon \simeq 10^{-6}$ which demonstrates that these winds
are Poynting flux driven, even if they would be ejected from a hot corona
with $p_* \simeq 10^{-6}$.

{\it The ratio $\epsilon/p_*$ is the true measure for the question of
Poynting flux driven outflows. For relativistic winds, $\epsilon \rightarrow 1$,
and the Poynting flux can then even exceed the rest mass energy flux. } 

\subsection{The wind equation}
In ideal MHD, plasma flows along the magnetic surfaces $\Psi = const$.
Since the Lorentz factor, the angular momentum (or angular velocity
$\Omega$) and the poloidal current function $I$ are essentially only
functions of the radius along the flux surface and the Mach number,
the normalisation of the plasma 4--velocity, $u_\alpha u^\alpha = -1$,
\begin{equation}
  \gamma^2 - \vec{u}^2 = 1 \quad,\quad \vec{u} = \gamma\vec{v}
\end{equation}
leads to the equation
\begin{equation}
  u_p^2 + 1 = \gamma^2 - \frac{j^2}{R^2} \,.
\end{equation}
 Using the solutions for $\gamma$
and $j$, this can be rearranged into the form
\begin{equation}
  \alpha^2 u_p^2 + \alpha^2 = \left( {E\over\mu} \right)^2\,
  V_{\rm S}(R; M^2,\epsilon) \,,
\end{equation}
where the subscript $S$ is for `Schwarzschild'. With radii measured in units of the light cylinder radii $R_L = c/\Omega^F$,
the potential function $V_{\rm S}$ can be expressed as
\begin{equation}
  V_{\rm S}(x;M^2,\epsilon) = {{F_{\rm S}(x;M^2,\epsilon)}\over{x^2D_S^2}}
\end{equation}
with the definitions
\begin{eqnarray}
  D_S &=& \alpha^2 - (R/R_L)^2 - M^2 \label{equ:ds}\,,\\
  F_S &=& x^2[\alpha^2(1-\epsilon) - M^2]^2 \nonumber\\
      &-& \alpha^2 [(1-\epsilon)x^2 - M^2\epsilon ]^2\,.
\end{eqnarray}
This relation is also correct near the Schwarzschild horizon, since $\alpha u_p$
remains finite there.

The above equation represents an algebraic relation for $M^2$ for each position
along a flux surface. This follows from the relation between $u_p$ and $B_p$
\begin{equation}
  (\alpha u_p)^2 = {{\eta^2B_p^2}\over n^2} = 
  {B_p^2\over{16\pi^2\mu^2\eta^2}}\,M^4 \,.
\end{equation}
Multiplying the equation by $x^4$ we obtain
\begin{equation}
  x^2\,{{F_S(x;M^2,\epsilon)}\over D_S^2(x;M^2)} \,
   \left( {E\over\mu} \right)^2
  = \alpha^2x^4 + {{B_p^2x^4}\over{16\pi^2\mu^2\eta^2}}\,M^4 \,.
\end{equation}
The last term can be written in terms of the flux function
\begin{equation}
  \Phi_\Psi(x) \equiv {{B_pR^2}\over{B_{p*}R_*^2}} \,,
\end{equation}
which is normalized by the footpoint $R_*$ and its magnetic field strength
$B_{p*}$. In addition, we introduce the {\it dimensionless magnetisation
parameter}
\begin{equation}
 \fbox{$ \displaystyle
  \sigma_*(\Psi) \equiv {{(B_{p*}R_*^2)(\Psi) c}\over{4\pi\mu\eta(\Psi)
  R_L^2(\Psi)}} \,,$}
\end{equation}
which is the analogue of the magnetisation parameter introduced in the
Newtonian discussion. This altogether leads to the equation
\begin{equation}
  \fbox{$ \displaystyle
  x^2\,{{F_S(x;M^2,\epsilon)}\over D_S^2(x;M^2)} \,
     \left( {E\over\mu} \right)^2
  = \alpha^2 x^4 + \sigma_*^2\,\Phi_\Psi^{2}\,M^4\,.$}
\end{equation}  
This fundamental equation clearly shows that the solutions of the wind
equation, when formulated for the Mach number $M$, depends on the following
parameters:
\begin{itemize}
\item the parameter $\epsilon$, which defines the position of the Alfv\'en 
point
\item the dimensionless energy $\bar E = E/mc^2$,
\item the magnetisation parameter $\sigma_*$,
\item the flux tube function function $\Phi_\Psi$
\item two additional parameters hidden in the specific enthalpy
\begin{equation}
  \mu =m c^2\Biggl[1 + {\Gamma\over{\Gamma -1}}\,p_*\,\left( {M_*^2\over M^2}
   \right)^{\Gamma - 1}\Biggr]\,.
\end{equation}
\end{itemize}
These are essentially 5 parameters for each flux surface $\Psi = const$.
It can be shown that the relativistic wind equation has the same critical
points as the non--relativistic one: the slow magnetosonic point,
the Alfv\'en point $D_S(R_A) = 0$ and the fast magnetosonic point. It
is very important in this regard that the light cylinder is not a critical
point \footnote{These points are modified somewhat when the structure of the magnetosphere is taken into account (Tsinganos et al. 1996).}.

{\it The requirement that a wind solution passes through all three
critical points fixes therefore three of the five parameters. We may consider
$\sigma_*$ and $p_*$ as free parameters which are fixed by injection physics.}

When pressure is as important as Poynting flux, $p_* \simeq \epsilon$, we
can apply the expansion of the wind equation
\beq
  x^2F_SE^2 = \bar\mu^2\,(D^2\alpha^2x^4 + \sigma_*^2\Phi^{2}D^2M^4)
\eeq
with $\bar\mu = \mu/mc^2$
\beq
  \bar\mu^2 = 1 + {{2\Gamma}\over{\Gamma - 1}}p_*\,\left( {M_*^2\over M^2}
    \right)^{\Gamma - 1} + O(p_*^2) \,.
\eeq
Since protostellar outflows are probably nearly isothermal (Paatz \& Camenzind 
1996), $\Gamma \simeq 1.2$, the energy equation can also be transformed into a
polynomial form for the square of the Mach number
\beq
  \sum_{n=0}^{n_p} m_{2n}\,M^{2n} = 0 \,,
\eeq
with a suitable power $n_p$ ($n_p = 9$ for $\Gamma = 1.2$). This form of the
wind equation can then be used to study all real branches (for a Newtonian
study, see Paatz \& Camenzind 1996).

\subsubsection{The cold wind equation}
In the cold limit, $p_* \ll \epsilon$, we may neglect the pressure terms in
the specific enthalpy, $\mu \simeq mc^2$, and the wind equation then provides a 4th degree polynomial in the square of the Mach number for the poloidal velocity
\beq
  \sum_{n=0}^4 m_{2n}\,M^{2n} = 0, \label{equ:pol}
\eeq  
with the following polynomial coefficients
\bea
  m_0 &=& E^2\alpha^2(1-\epsilon)^2 x^4(\alpha^2 - x^2) - 
    \alpha^2x^4(\alpha^2 - x^2)^2\, , \label{equ:79} \\
  m_2 &=& - 2E^2\alpha^2(1-\epsilon)^2x^4   + 2\alpha^2x^4(\alpha^2 - x^2)\, , \\
  m_4 &=& E^2x^2(x^2 - \alpha^2\epsilon^2) - \alpha^2x^4 - \sigma_*^2 \Phi^{2}
    (\alpha^2 - x^2)^2\, , \\
  m_6 &=& 2\sigma_*^2\Phi^{2}(\alpha^2 - x^2)\, , \\
  m_8 &=& - \sigma_*^2\Phi^{2}  \,,
\eea 
and with $m_{2n}=m_{2n}(E,\epsilon,\sigma_*,\Phi;x)$.
Without gravity ($\alpha = 1$), these expressions are identical with the flat
space expansion (Camenzind 1986b). In the case of a cold wind $u_{p}(x_{\mathrm{inj}})=0$, where $x_{\mathrm{inj}}$ is the injection point of the plasma from the star, and $m_{0}(x_{\mathrm{inj}})=0$. Consequently, we get a relation between injection radius, energy and Alfv\'{e}n radius
\begin{equation}
\epsilon=x_{A}^2=1-{\frac{\sqrt{\alpha^2-x_{\mathrm{inj}}^2}}{E\,\alpha}},\label{equ:80}
\end{equation}
hence  $m_{2n}=m_{2n}(x_{inj},\sigma_*,\Phi;x)$. A cold wind has only two critical points, the fast magnetosonic one $x_{FM}$, where the wind velocity must equal the fast magnetosonic velocity, and the Alfv\'en point. The slow magnetosonic point $x_{\mathrm{SM}}$ disappears, with the consequence that the mass flux or the magnetisation along a flux surface becomes a free parameter. The solution of Eq. (\ref{equ:80}) is uniquely determined by the critical energy $E_{\mathrm{crit}}(\Psi)$. 
Eq. (\ref{equ:pol}) is a 4th order polynomial. For this reason the wind equation has 4 solutions $u_{p,i}(x),(i=1,...,4)$ at each radius $x$. We consider negative or complex solutions to be 'unphysical'. We obtain two positive branches of which the `physical' solution is the one which passes through the critical points and starts with zero poloidal velocity. In order to determine the critical solution, we use a bracketing method varying the energy parameter $E(\Psi)$ until the critical solution is found. For energies $E<E_{\mathrm{crit}}$, the wind ceases before being able to reach the fast magnetosonic point or the asymptotic radius, hence there is no continuous wind solution for all $x$. In the case  $E>E_{\mathrm{crit}}$, the two branches of the solution do not meet at the fast magnetosonic point (see also Fendt \& Camenzind 1996) and no wind solution is found either. We fix the free parameter $\epsilon$  by choosing the values for the injection radius $x_{\mathrm{inj}}$ and the magnetisation parameter $\sigma_*$. For small $\sigma_*$ the energy must be determined up to the order $10^{-2}\sigma_*$.

\section{The transfield equation and its action integral}\label{chap:3}
Amp\`ere's equation determines the form of the magnetic flux surfaces, known
as the Grad--Schl\"uter--Shafranov equation for $\Psi$
\beq
  \Delta^*\Psi \equiv {R^2\over\alpha}\,
   \nabla\cdot\left( {\alpha\over R^2}\nabla\Psi \right) =
    - {{4\pi}\over c}\,Rj_\phi\,.\label{equ:81}
\eeq    
$\nabla$ is here the covariant derivative with respect to 3--space. This
divergence--operator can explicitly be written as
\beq
  \Delta^* = {\partial\over{\partial r}}\left( \alpha^2\,{\partial
    \over{\partial r}} \right) +
    {\sin\theta\over r^2}\,{\partial\over{\partial\theta}}
    \left( {1\over\sin\theta}\,{\partial\over{\partial\theta}} \right) \,
\eeq    
and is equivalent to the flat space analog except for the redshift factor $\alpha$.

The split--monopole
\beq
  \Psi = \Psi_*(1 - \cos\theta)
\eeq
is still a solution of the vacuum equation. A dipole
\beq
  \Psi = \Psi_* \frac{\sin^2\theta}{r}
\eeq
would be modified somewhat by the gravitational potential.    

\subsection{The Grad--Schl\"uter--Shafranov equation}
When the magnetosphere is filled with plasma, currents will deform the vacuum
solutions.
The current density $j_\phi$ follows from the transverse component of the
equation of motion 
\beq
  \epsilon\nabla_\perp\alpha + \nabla\cdot(\alpha\tensor{t})_\perp = 0 \,. 
\eeq 
The explicit form of this equation has been derived in various papers
(Mobarry \& Lovelace 1986; Camenzind 1987; Beskin et al.
1995; Fendt et al. 1995) 
\beq
  R^2\nabla\cdot\left( {{\alpha D_S}\over{\alpha^2R^2}}\nabla\Psi \right) =
  g(I,M^2,...)\label{GSS}
\eeq
with $D_S$ as defined in Eq. (\ref{equ:ds}) and $g(I,M^2,...)$ a source function. 

This equation has various interesting limits. One is the force--free limit
with vanishing pressure and $M^2 = 0$ (Blandford \& Znajek 1976; Mobarry \&
Lovelace 1986; Okamoto 1992; Fendt 1997)
\beq
  R^2\nabla\cdot\left( {D_S\over{\alpha R^2}}\nabla\Psi \right) +
    {2\over\alpha}\,{{dI^2}\over{d\Psi}} = 0 \,.
    \label{ff}
\eeq
In this case, $D_S = \alpha^2 - x^2$ and the source of the magnetic flux
is only given by the poloidal current $I(\Psi)$. Even in this limit, the
equation is highly non--linear and the Alfv\'en surface now coincides with
the light cylinder. Such solutions are interesting for black hole driven
outflows.

Another limit which has been discussed quite often in the literature is the
Newtonian limit without gravity, $\alpha = 1$ and $R \ll R_L$
(Heinemann \& Olbert 1978;
Pelletier \& Pudritz 1992; 
Rosso \& Pelletier 1994)
\beq
  \nabla\cdot\left( {{M^2 - 1}\over R^2}\,\nabla\Psi \right) =g_N(I,M^2,...).
\eeq  
Rosso \& Pelletier (1994) have used this limit for the
discussion of disk magnetospheres. They have derived a Lagrangian description
of the form
\beq
  \nabla\cdot\left( {{M^2 - 1}\over R^2}\,\nabla\Psi \right) -
    {{\partial U_0}\over{\partial\Psi}} = 0 
\eeq  
depending on a potential function $U_0(M^2, \Psi;R,z)$. It is then evident that
this equation can be derived from the Lagrangian
\beq
  L_0(M^2,\Psi; R,z) = {1\over 2}\,{{M^2 - 1}\over R^2}\,(\nabla\Psi)^2
    + U_0(M^2,\Psi;R,z)
\eeq    
and the proper action of the system is simply
\beq	
  {\cal S}[\Psi] = \int 2\pi\,L_0(M^2,\Psi;R,z)\,R\,dR\,dz \,.
\eeq  
The above formulation is indeed only a special case of the action discussed
by Mobarry \& Lovelace (1986).
\subsection{On the causal structure of self--consistent magnetospheres}
As shown above, the concept of light cylinders appears naturally in ideal relativistic MHD. This condition is largely satisfied in thin outflowing plasmas, in particular when plasma will be reheated by some processes (shocks e.g.). The light cylinder is not a critical surface of the Grad--Schl\"uter--Shafranov equation (see e.g. Beskin \& Pariev 1993) except in the force--free limit. As in the Newtonian case, the relativistic Grad--Schl\"uter--Shafranov equation has the same critical surfaces, the slow and the fast magnetosonic surface, as well as the Alfv\'en surface (Heinemann \& Olbert 1978, Tsinganos et al. 1996). The system becomes hyperbolic beyond the fast magnetosonic surface, so that no back--reaction with the star is allowed from this region. 
\section{Collimation of spherical outflows}\label{chap:4}
The collimation of magnetized winds into a conical outflow is very similar
to the plasma confinement in a $z$--pinch. In this region we can neglect
gravity. The equilibrium condition for a non--rotating axisymmetric current 
carrying pinch configuration (Freidberg 1987) 
\beq
  \kappa\,{B_p^2\over{4\pi}} = \nabla_\perp
  {B_p^2\over{8\pi}} + \nabla_\perp {B_\phi^2\over{8\pi}} + \nabla_\perp P
  - {I^2\over{4\pi R^3}}\,(-\nabla_\perp R)\label{equ:89} 
\eeq
has to be 
generalized to include various relativistic effects (Appl \& Camenzind 1993a,b)
\begin{eqnarray}
  \kappa\,{B_p^2\over{4\pi}}\,(1 - M^2 - x^2) &=& (1-x^2)\,\nabla_\perp
  {B_p^2\over{8\pi}} + \nabla_\perp {B_\phi^2\over{8\pi}} + \nabla_\perp P
  \nonumber \\  
  - {{B_p^2\Omega^F}\over{4\pi c^2}}\,\nabla_\perp (\Omega^FR^2) &+&
  \left( {{\mu nj^2}\over R^3} - {I^2\over{4\pi R^3}} \right)\,
  (-\nabla_\perp R) \,.\label{fbal}
\end{eqnarray}
Pressure gradients acting perpendicular to the magnetic surfaces are in
equilibrium with electric forces, centrifugal forces, pinch forces produced 
by poloidal currents, as well as with  curvature forces expressed at the
left hand side. This latter force is modified by the Mach number $M$ and 
the light cylinder, and $\mu = (\rho + P)/n$. A comparison between the Newtonian expansion (\ref{equ:89}) and the special relativistic expression (\ref{fbal}) clearly demonstrates the influence of the light cylinder in the force--balance equation. The major distinction is found in the first expression on the right--hand side, $(1-x^2)\,\nabla_\perp B_p^2/8\pi$. This term changes its sign at the position of the light cylinder, $x=1$, as a consequence of the existence of electric fields in Maxwell's stress tensor $\tensor{t}_{em}$, Eq. ({\ref{equ:25}). Since the observed collimation radii are larger than the light cylinder, field lines anchored in the star will cross the light cylinder, at least when the stellar rotation periods are of the order of a few days.

The specific angular momentum of the plasma $j = R^2\Omega =R u_\phi/\gamma$ 
is now given by the wind theory
\beq
  j = R_Lc\,{{(1-\epsilon)x^2 - M^2\epsilon}\over {1 - \epsilon - M^2}}
\eeq
and depends on the Mach number $M$. Outside the Alfv\'en surface, $M^2 > 1$,
this can be expressed as
\beq
  j \simeq R_Lc\,\left( \epsilon - {{(1-\epsilon)x^2}\over M^2} \right) \,.
\eeq
Outside the Alfv\'en surface, the specific angular momentum is essentially
constant and has the value
\beq
 \fbox{$ \displaystyle
  j \simeq \epsilon\,R_Lc = R_A^2\Omega^F \,. $}
\eeq

Similarly, the pinch current $I$ is not
a free quantity, but also follows from the wind theory
\beq
  I = {{2\pi\eta E}\over\Omega^F}\,{{\epsilon - x^2}\over
    {1 - x^2 - M^2}}\,. 
\eeq
Using the expression for the Lorentz factor, the total energy $E$ can be replaced
by, $\alpha = 1$,
\beq
  I = {{2\pi\eta\mu}\over{\gamma\Omega^F}}\,{{\epsilon - x^2}\over
    {1 - \epsilon - M^2}}\,. 
\eeq
For radii $x^2 > \epsilon$ and Mach numbers $M^2 > 1$ , i.e. essentially 
outside the Alfv\'en surface, the current can be approximated by
\beq
  I \simeq {{2\pi\mu\eta}\over{\gamma\Omega^F}}\, {x^2\over M^2} =
    {{2\pi\mu\eta}\over{\gamma\Omega^F}}\, {x^2\over{4\pi\mu\eta}}\,
    {n\over\eta} \,.
\eeq
With the definition of $\eta$ this gives the universal expression for the
current
\beq
 \fbox{$ \displaystyle
  I \simeq {1\over 2}\,\Omega^F(B_pR^2)\,{1\over\beta_p} $}
\eeq
always valid outside the Alfv\'en surface. $\beta_p = v_p/c$ is the
dimensionless poloidal velocity of the plasma.   

\subsection{Asymptotic collimation}
We discuss in the following the solutions for a cylindrical pinch,
where the curvature forces vanish, $\kappa = 0$. In this case, the equilibrium
condition reduces to its one--dimensional form
\bea
 (1-x^2)\,{d\over{dR}}\,
  {B_p^2\over{8\pi}} &+& {d\over{dR}} {B_\phi^2\over{8\pi}} + {{dP}\over{dR}}
    - {{B_p^2R}\over{2\pi R_L^2}} \nonumber \\
  &-& \left( {{\mu nj^2}\over R^3} - {I^2\over{4\pi R^3}} \right) = 0
     \,.\label{f1bal}
\eea 
The pressure of the toroidal field and the pinch force can be combined into
one single expression. In addition, we normalize the 
poloidal magnetic field in units of its value on the central axis, $B_0$,
$y = B_p^2/B_0^2$,
\begin{equation}
  (1 - x^2)\,{{dy}\over{dx}} - 4xy + {1\over x^2}\,{{dI^2}\over{dx}}
   + {{8\pi}\over B_0^2}\,{{dP}\over{dx}} - 
   {{8\pi\mu n}\over{R_L^2B_0^2}}\,{j^2\over x^3} = 0\,.
   \label{pinch}
\end{equation} 
Electric fields, $E_\perp = (R/R_L)B_p$, due to the rotation of the magnetic 
surfaces modify the classical pinch equilibrium in two ways. First,
the equation has a singular point at the light cylinder, and secondly,
the additional term $4xy$ is crucial for the form of the equilibrium. In 
contrast to force--free models (Appl \& Camenzind 1993b), the specific angular 
momentum $j$ of the plasma and the poloidal current $I$ are not arbitrary, 
but follow from the MHD wind theory.

In particular, the last term can be written as
\beq
  {{8\pi\mu n}\over{R_L^2B_0^2}}\,{j^2\over x^3} =
   2\,{B_p^2\over B_0^2}\,{v_{\phi,A}^2\over V_A^2}\,{R_A^2\over R_L^2}\,
   {\bar j^2\over x^3} = 2y\,{R_A^2\over R_L^2}\,{\bar V}(x)\,
   {\bar j^2\over x^3}\,,
\eeq
where $V_A$ is the local Alfv\'en speed and $v_{\phi,A} = R_A\Omega^F$ the
rotational velocity at the Alfv\'en point. We also expressed the specific
angular momentum in terms of its characteristic value, 
$j = \bar j\,R_A^2\Omega^F$. This term contributes therefore
to the second term in the pinch equation (\ref{pinch}). This shows therefore
that the centrifugal force is not important near the light cylinder and
beyond, whenever the Alfv\'en radius is far inside the light cylinder. In this
domain the electric fields provide the important forces. Far inside the
light cylinder, however, centrifugal forces are important.

\subsection{Pinch solutions without angular momentum}   
We first discuss solutions of the relativistic pinch equilibrium neglecting
the influence of the centrifugal force. In this case the equilibrium is 
determined by the condition  
\begin{equation}
  (1 - x^2)\,{{dy}\over{dx}} - 4xy + {1\over x^2}\,{{dI^2}\over{dx}}
   + {{8\pi}\over B_0^2}\,{{dP}\over{dx}} = 0\,.
\end{equation} 
The toroidal current follows to a good approximation
\begin{equation}
  I \simeq - \Omega^F(R^2B_p)\,{1\over\beta_p} = -cB_0R_L\,(x^2\sqrt{y})
       \,{1\over\beta_p}
\end{equation}  
and we prescribe the plasma pressure in terms of $P = \beta B_p^2/8\pi$
with a constant plasma beta.
This leads finally to the equation, $\gamma_p = 1/\sqrt{1 - \beta_p^2}$,
\begin{equation}
  \left( 1 + \beta + {x^2\over{\gamma_p^2\beta_p^2}} \right)\,{{dy}\over{dx}}
    + {4\over{\gamma_p^2\beta_p^2}}\,xy = 0\,.
\end{equation}
This homogeneous equation has the remarkably simple solution
\begin{equation}
  y(x) = {C\over{(1 + \beta + x^2/x_c^2)^2}}
\end{equation}
with 
\begin{equation}
  x_c = \gamma_p\,\beta_p \simeq 10^{-3} \quad,\quad R_c = \gamma_p\beta_p\,R_L
    \simeq 10^{-3}\,R_L 
\end{equation}
as the {\em core radius} of the jet with $\beta_p \simeq 10^{-3}$ and
$\gamma_p = 1$ for protostellar outflows. Inside the core, given by the light 
cylinder radius and 
the poloidal velocity $\beta_p$, the poloidal magnetic field is
essentially homogeneous. It decays as a monopole field outside the core,
$B_p \propto 1/R^2$ for $R \gg R_c$
\begin{equation}
  B_p = {B_0\over{1 + \beta + R^2/R_c^2}}\,.
\end{equation}
This is not unexpected since the jet is formed by a supermagnetosonic wind.
The current function $I$ increases quadratically inside the core and saturates
towards a constant value outside the core, $I_\infty \simeq cB_0R_L\beta_p$.
This is exactly the structure of the force--free solutions found previously
(Appl \& Camenzind 1993b). 

The flux function has the same form as in the force--free solutions
\beq
  \Psi \propto \ln[1 + \beta + (R/R_c)^2 ] \,.
\eeq
This is not unexpected, since in the collimated regime the magnetic structure
is force--free, $I = I(\Psi)$.
  
{\it MHD jets always have a core--envelope structure, along the
axis the magnetic field is predominantly longitudinal, beyond the core
it becomes dominated by the toroidal component. The light cylinder
$R_L \simeq 100$ AU is the basic scale for the jets, the core is
extremely small, $R_c \simeq 10R_*$.}    

\subsection{Pinch solutions including angular momentum}
In this section we investigate the behaviour of the pinch equation inside
the light cylinder including centrifugal forces
\bea
  (1 - x^2)\,{{dy}\over{dx}} - 
    \left(4x + C\,{{\bar V(x)\bar j^2}\over x^3}\right)\,y & & \nonumber\\
    + {1\over x^2}\,{{dI^2}\over{dx}} + {{8\pi}\over B_0^2}\,{{dP}\over{dx}} & = & 0 \,.
\eea
The centrifugal force must be regularized inside the Alfv\'en surface. A
suitable function which guarantees corotation inside the Alfv\'en point is
given by
\beq
  \bar j = {{(R/R_A)^2}\over{1 + (R/R_A)^2}}\,.
\eeq   
Using the expression for the current and neglecting the pressure
\begin{equation}
  \left( 1 + {x^2\over{\gamma_p^2\beta_p^2}} \right)\,{{dy}\over{dx}}
    + \left( {4\over{\gamma_p^2\beta_p^2}}\,x + 
      C\,{{\bar V(x)\bar j^2}\over x^3}\right)\,y 
    = 0\,,
\end{equation}
with solutions as obtained by a simple integration
\beq
 \ln y = \int {{ (4/\gamma_p^2\beta_p^2)\,x + 
    C\,{\bar V(x) \bar j^2}/x^3 }\over
   {1 + x^2/(\gamma_p^2\beta_p^2) }}\,dx \,,
\eeq 
or
\beq
  \ln y = \ln\left( {1\over{1 + x^2/x_c^2}} \right)^2 +
    \int {{C\,\bar V(x) \bar j^2}\over
    {x^3[1 + x^2/x_c^2] }}\,dx \,.
\eeq      
The second integral only contributes in the region of the Alfv\'en radius.
For the case $R_c = R_A$ and $\bar V = const$, the solution is simply given by
\beq
  \ln y = \ln\left( {1\over{1 + x^2/x_c^2}} \right)^2 -
    {C\over 4}\,{R_L^2\over R_A^2}\,{1\over{(1 + R^2/R_A^2)^2}} + C_1 \,.
\eeq      
Since $C \propto R_A^2/R_L^2$, angular momentum modifies the solution only
near the Alfv\'en surface. These effects drop out very rapidly beyond the
Alfv\'en radius, and we are left with the solution found previously.

\section{A model for collimated outflows}\label{chap:5}
Fully self--consistent solutions of the Grad--Shafranov equation are not yet
available. We want to demonstrate the main features of collimated outflows
in terms of a model which satisfies the inner boundary conditions and
the asymptotic flux distribution for cylindrical outflows. The existence of
a disk surrounding the star will modify the dipolar field in the innermost
few stellar radii (see Fendt et al. 1995). Beyond the inner edge of the disk
the resulting magnetosphere is however largely spherical and cylindrical in the asymptotic region.

\subsection{The model}
The flux distribution should be a solution of the Grad-Shafranov equation. As a first
approximation we use the following ansatz for the flux function, where $\Psi$ consists of a homogeneous and a special solution
\begin{equation}
\Psi(r,\theta)=\Psi^{\mathrm{hom}}(r,\theta)+\Psi^{\mathrm{spec}}(r,\theta),\label{psitot}
\end{equation}
with
\bea
\Delta^*\Psi^{\mathrm{hom}}& = & 0, \nonumber \\
\Delta^*\Psi^{\mathrm{spec}}& \propto & j_{\phi}.\nonumber \\
\eea
The vacuum solution (see Appendix A)  is given by
\bea
\Psi^{\mathrm{hom}}(\rho,\theta) & = & \frac{2\mu_*\,\rho}{\pi\,a_{\mathrm{disk}}}\left(\frac{\psi}{\rho}\right.+\gamma(1-\psi)
-\frac{\cos^2\theta}{\rho^2\gamma}\nonumber\\
 & & +\frac{\sin^2\theta}{\rho^2}\left.\left(-\arctan\frac{1}{\gamma}+\frac{\pi}{2}\right)\right),\label{equ:7}
\eea
where
\begin{equation}
\gamma:=\frac {\sqrt{2}\,|\cos\theta|}{\sqrt{\rho^{2}-1+\sqrt{(\rho 
^{2}-1)^{2}+4\rho^{2}\cos^2\theta}}}.
\end{equation}
The quantity $\rho$ is dimensionless due to the normalization of the spherical radius $r$ in units of the inner radius of the disk $a_{\mathrm{disk}}$, $\rho:=r/a_{\mathrm{disk}}$. We assume the existence of a gap between the star and the disk due to the magnetic pressure. The dipole moment is given by $\mu_*=R_{*}^3B_{*}$. For our calculations the stellar radius is $R_{*}=3R_{\odot}$ and the maximum magnetic field strength on the stellar surface $B_{*}=1000$ G. The last term in $\Psi^{\mathrm{hom}}$ denotes the dipole magnetic flux generated by the star itself, while the remaining terms denote the flux contribution from the screening of the disk (\cite{kundt}); \cite{aly}; \cite{paatz96}). The integration constant $\psi$ equals the flux on the open field lines for the homogeneous solution. For an integration constant $\psi=0$ all flux is forced to go through the hole between star and disk and cannot escape to infinity (\cite{kundt}). The term $\psi/\rho$ does not contribute to the magnetic field, but guarantees $\Psi=0$ on the z-axis.  
In the collimation region we find for large radii from Eq. (\ref{equ:7})
\begin{eqnarray}
\gamma&=&\frac{\cos\theta}{\rho}+{\cal O}(\frac{1}{\rho^3}), \\
\Psi^{\mathrm{hom}}(\rho,\theta)&=&\frac{2\mu_*}{\pi\,a_{\mathrm{disk}}}\psi(1-\cos\theta)+{\cal O}(\frac{1}{\rho^2}).\label{equ:136}
\end{eqnarray}
This means that $\Psi^{\mathrm{hom}}$ behaves like a monopolar type function for large radii.

$\Psi^{\mathrm{spec}}$ describes the magnetic flux produced by a toroidal current $j_{\phi}$
\begin{equation}
\Psi^{\mathrm{spec}}(\rho,\theta)=A\, \ln\left(1+\frac{a^2}{R_{c}^2}(1-\cos\theta)+\frac{r^2\sin^2\theta}{R_{c}^2}\right). \label{equ:10}
\end{equation}
Apart from the normalisation factor $A$, which will be discussed in Sect. \ref{sec:weight}, it contains two essential radii. $R_c$ is
the core radius discussed in the last section, since for large radii we
find then
\beq
  \Psi^{\mathrm{spec}} \propto \ln\Bigl[ 1 + {{r^2\sin^2\theta}\over R_c^2} \Bigr]\label{equ:137}
\eeq
as required by cylindrical collimation. For small radii, the cylindrical radius
drops out of the expansion and we get a monopolar type solution
\beq
  \Psi^{\mathrm{spec}} \propto \ln\Bigl[ \left( {a\over R_c} \right)^2\,(1 - \cos\theta)
    \Bigr] \, \label{equ:12}.
\eeq
The parameter $a$ is a kind of jet radius, since the last magnetic surface
that can be closed with the star is given by $\ln(a/R_c)^2$. In the case of Eq. (\ref{equ:10}),
the flux surfaces can be parameterized by means of
\beq
  r(\theta) = R_c\,{\sqrt{ \tilde{C} - 1 -a^2(1 - \cos\theta)/R_c^2} \over\sin\theta}\,.
\eeq  
The angle $\theta$ extends from $\theta = 0$ (asymptotic region) to a maximal
value $\theta_{\tilde{C}}$, where the flux surface intersects with the stellar surface,
$r(\theta_{\tilde{C}}) = R_*$. The flux surface label $\tilde{C}$ is therefore given by this angle
\beq
  \tilde{C} = 1 + {R_*^2\over R_c^2}\,\sin^2\theta_{\tilde{C}} + {a^2\over R_c^2}\,
    (1 - \cos\theta_{\tilde{C}}) \,.
\eeq 

For the total flux function the collimation part will finally always dominate the homogeneous part. Comparing  Eq. (\ref{equ:136}) with Eq. (\ref{equ:137}) we see that the first does not depend on the radius, hence the latter containing the square of the radius will determine the solution for sufficiently large radii.

Note furthermore that we need all parts of Eq. (\ref{psitot}) to reproduce the dipole influence of the star at small radii and the collimation of the field lines for the asymptotic radii. Excluding the monopolar type part of Eq. (\ref{equ:10}) we cannot obtain realistic jet radii and the magnetosphere will collimate much too quickly.
$\Psi^{\mathrm{spec}}$ alone cannot reproduce the influence of the dipole near the star. The surfaces of constant flux for the total flux function Eq. (\ref{psitot}) are calculated with the Newton-Raphson method. The resulting magnetic fields can be found in Appendix B.

\subsection{The amount of flux through the hole of the disk}\label{chap:52}
The magnetic flux  $\Psi_{\mathrm{hole}}$ through the hole of the disk at $\mathrm{z}=0$ consists of the flux belonging to the disk field $\Psi^{\mathrm{hom}}_{\mathrm{disk}}$ and $\Psi^{\mathrm{spec}}$ between star and disk, 
the whole flux of the dipole $\Psi_{\mathrm{Dip}}$ minus the part of the dipole on the disk
(\cite{riffert})
\begin{eqnarray}
\Psi_{\mathrm{hole}} &=&\frac{2\mu_*\,\rho}{\pi\,a_{\mathrm{disk}}}
 \left(\gamma(1-\psi)
-\frac{\cos^2\theta}{\rho^2\gamma}-\right.\nonumber \\ 
& & \hspace{1.3cm}\left. \left.\frac{\sin^2\theta}{\rho^2}
\arctan \frac{1}{\gamma}\right)
\right]_{\rho=0}^{1} 
 \nonumber \\
 &   &+ \left. \Psi^{\mathrm{spec}}\right]
_{\rho=0}^{1}+\Psi_{\mathrm{Dip}}-\left.\frac{\mu_*\,\sin^2\theta}{r}\right]_{\rho=1}^{\infty}.
\end{eqnarray}

For $\psi=0$ all other terms except for the flux of the dipole $\Psi_{\mathrm{Dip}}$ and $\Psi^{\mathrm{spec}}$ will vanish (see also \cite{riffert}). Then plasma can only escape to infinity due to the special solution. This characteristic is discussed in Sects. \ref{chap:63} and  \ref{chap:64}. For $\psi\neq 0$ and $A=0$ the amount of flux through the hole is determined by the integration constant $\psi$ and the disk radius $a_{\mathrm{disk}}$ via $\Psi_{\mathrm{hole}}\propto -\psi/a_{\mathrm{disk}}$. This means that for $\psi=0$ and $A=0$ all flux goes through the gap between the star and the disk. This amount will decrease with increasing $\psi$ and $A$.

\section{Results and discussion}\label{chap:6}
We now present solutions for the magnetospheres and the resulting winds which are calculated along the collimating jet flux surfaces. This leads us to a 2D velocity distribution. We discuss the impact of the integration constant $\psi$, the disk radius $a_{\mathrm{disk}}$, the normalisation factor $A$ of $\Psi^{\mathrm{spec}}$ and the redshift factor $\alpha$ on the wind solutions.
Table \ref{tab:1} summarizes the values of all other parameters which are kept constant throughout this paper.
\begin{figure*}
\begin{center}
\resizebox{\hsize}{!}{\includegraphics{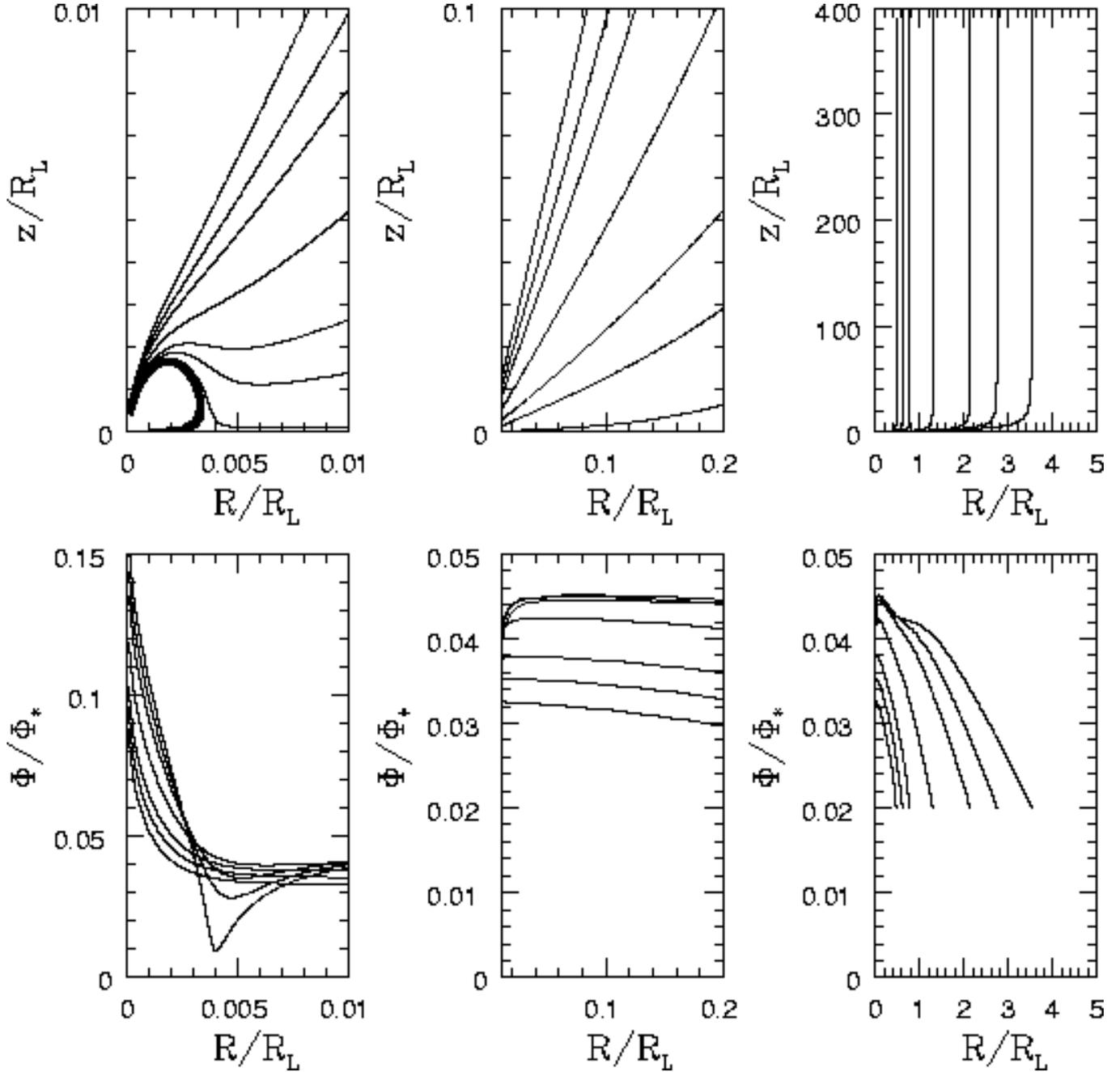}}
\caption{Top: Magnetic flux surfaces $\Psi$ for $a_{\mathrm{disk}}=2.5R_*$, $A=10$ and $\psi=0.1364$. The contour levels are \mbox{$\Psi=1.01, 1.00, 0.96, 0.93, 0.86, 0.78, 0.75, 0.71$}. The thick curve in the left panel represents the last closed flux surface and marks the beginning of the disk at $a_{\mathrm{disk}}$.
Bottom: The corresponding flux tube functions $\Phi$}
\label{fig:1}
\end{center}
\end{figure*}

\begin{table}[h]  
\caption{{\small Model parameters}}
\label{tab:1}
\begin{center}
\begin{tabular}{|l|l|}
\hline
 & \\[-.3cm]
Model parameter & Value  \\
\hline
 & \\[-.3cm]
Stellar radius $R_*$ & 3$R_{\odot}$\\
Light cylinder $R_L$ & 2250$R_*$ \\
Injection radius $r_{\mathrm{inj}}$ & 1$R_*$\\
Core radius $R_c$ & 8$R_*$\\
Jet radius $a$ of Eq. (\ref{equ:10}) & 1430$R_*$\\
Peak stellar magnetic field $B_*$ & 1000 G\\
Dipole moment $\mu_*$ & $R_*^3B_*$ \\
Stellar mass $M_*$ & $2M_{\odot}$\\
Stellar density $n_*$ & $10^{13}$cm$^{-3}$\\
Stellar Alfv\'en velocity $v_{A,*}$ & 690 km s$^{-1}$\\
Magnetisation $\sigma_*$ & $10^{-8}$ \\
\hline
\end{tabular}
\end{center}
\end{table}

The distribution of the magnetic flux surfaces $\Psi(R,z)$ and the corresponding magnetic flux tube functions $\Phi(R)$ for each surface are shown in Fig.~\ref{fig:1}. 
The critical surface $\Psi_{\mathrm{crit}}$ is the first open flux surface and reaches the asymptotic jet radius $R_{\mathrm{jet}}$. There is no jet at radii larger than that given by $\Psi_{\mathrm{crit}}$. We choose our parameter $a$ of Eq. (\ref{equ:10}) such that we get realistic jet radii of $R_{\mathrm{jet}}\approx (3-4)R_{L}$ (\cite{fendt96}). In our calculation the period of the star is $P_*=1.14$ days which is equivalent to a light cylinder at $R_{L}=2250R_{*}$. Because of the influence of $\Psi^{\mathrm{hom}}$ the jet radius is shifted to bigger values than those given by $a$. We get $R_{\mathrm{jet}}=7937R_{*}=3.5 R_{L}$ for $a=1430R_*$. The core radius $R_{c}=8R_{*}$.
As injection radius for the plasma we choose $r_{\mathrm{inj}}=1R_{*}$.

In contrast to calculations done by Fendt \& Camenzind (1996) our analytical model is not limited by numerics, hence we can take a realistic weak magnetisation parameter of $\sigma_*=10^{-8}$ which corresponds to observations. Throughout the jet $\sigma_*$ is kept constant which is equivalent to a constant mass flow rate on each surface $\Psi$. We normalize the radius $R$ to the light cylinder $R_{L}$ and the velocities to the Alfv\'en velocity on the stellar surface $v_{A,*}$, where
$v_{A,*}^{2}=B_{*}^{2}/4\pi m_{\mathrm{p}} n_{*}$, and the flux surfaces to 
$\Psi_{\mathrm{crit}}$. The density on the stellar surface is $n_{*}=10^{13}$cm$^{-3}$, corresponding to an outflow rate of $\dot M_{\mathrm{jet}}\simeq 10^{-8}$ M$_{\odot}$yr$^{-1}$. 

The flux tube function is a measure for the opening angle of the flux surfaces. Small $\Phi$ correspond to large openings, constant values to a conical structure of $\Psi$ ($\Phi=\mathrm{const}$ represents a monopolar type flux surface).
For small radii we can reproduce a stellar dipole which is mainly responsible for the strong decrease of the flux tube functions. The more the surfaces show a dipolar structure instead of a conical one, the longer is the extent of the region of decrease of the flux tube function. This is equivalent to an acceleration of the plasma. There is a deceleration of the plasma for the outer flux surfaces for radii larger than the disk radius. The corresponding $\Phi$-values then even catch up with those obtained by the more inward lying surfaces. Nevertheless, their total plasma acceleration always exceeds that of the latter ones. The deceleration process coincides with a sharp bending of these flux surfaces at the inner edge of the disk which is not found for the inner flux surfaces. From this we draw the conclusion that the inner dipole part of the star in our model is crucial for the plasma acceleration. 
The closer the flux surface is to the z-axis, the less the surfaces look dipole-like and show a conical structure instead. For intermediate radii the monopolar-type part of $\Psi^{\mathrm{spec}}$ controls all flux surfaces, consequently all flux tube functions become constant. 
For large radii all surfaces collimate.

\subsection{The cold wind}
A physical cold wind solution has to start at zero velocity at the injection radius and, accelerating monotonically, pass through the Alfv\'en point and the fast magnetosonic point. Figure \ref{fig:1a} shows the wind solution along the flux surface $\Psi=0.89$.
The maximum poloidal velocity is reached when $\Phi$ is lowest.
\begin{figure}
\resizebox{\hsize}{!}{\includegraphics{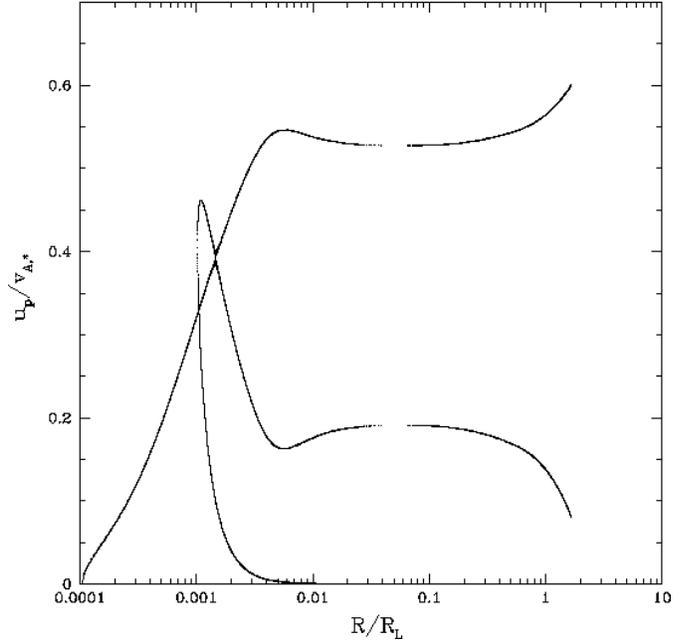}}
\caption{Poloidal wind velocity $u_p$ along the flux surface \mbox{$\Psi=0.89$} and for $a_{\mathrm{disk}}=2.5R_*$, $\psi=0.1364$. The physical branch starting with nearly zero velocity and passing through the Alfv\'en and fast magnetosonic points is shown together with the unphysical branch which is retained in the subsequent figures}
\label{fig:1a}
\end{figure}
 As a general feature it can be seen that the acceleration and deceleration of the plasma along different flux surfaces varies. The acceleration for the outer surfaces is more efficient than for the inner ones and the asymptotic poloidal velocity $u_p^{\infty}$ is larger. Figure \ref{fig:13} shows the asymptotic velocity distribution over the cross-section of the jet. The velocity decreases for decreasing flux surfaces. In our calculations, the mass flow rate per flux surface is conserved, $\eta(\Psi)=\mathrm{const}$.
\begin{figure}
\resizebox{\hsize}{!}{\includegraphics{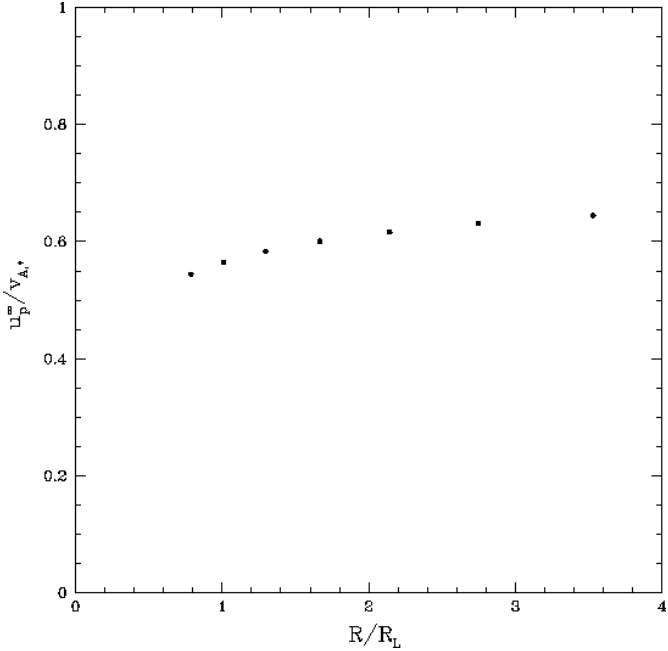}}
\caption{Asymptotic velocities $u_p^{\infty}$ for the flux surfaces \mbox{$\Psi=1, 0.96, 0.93, 0.89, 0.86, 0.82$} and $0.78$ and $a_{\mathrm{disk}}=2.5R_*$, $\psi=0.1364$}
\label{fig:13}
\end{figure}

The positions of the Alfv\'en point $x_A$ and the fast magnetosonic point $x_{FM}$ are plotted for several flux surfaces in Fig.~\ref{fig:3}. The resulting value for $\epsilon$ is of the order of $10^{-6}$.
\begin{figure}
\resizebox{\hsize}{!}{\includegraphics{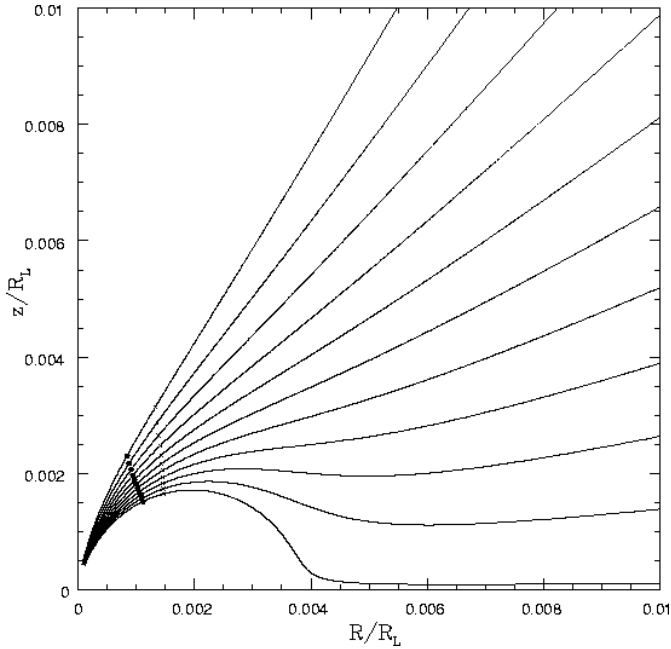}}
\caption{Magnetic flux surfaces $\Psi$ for $a_{\mathrm{disk}}=2.5R_*$ and $\psi=0.1364$. The dots represent the position of the Alfv\'en point, the crosses that of the fast magnetosonic point. The contour levels are $\Psi=1.0$ to $0.64$ in steps of $0.036$}
\label{fig:3}
\end{figure}
The Alfv\'en point moves to smaller radii for decreasing flux surface values $\Psi$. This is equivalent to a smaller amount of angular momentum being transferred to the plasma. As expected for weakly magnetized flows (\cite{fendt96}), the Alfv\'en point is very close to the injection radius and far from the light cylinder.
For successively decreasing surfaces the fast magnetosonic points move linearly inward towards smaller radii. 
For the innermost flux surfaces no determinations of $x_{FM}$ and $E_{\mathrm{crit}}$ are possible any longer. This means that for these surfaces the wind remains submagnetosonic.  Wind solutions for flux surfaces lower than $50\%$ of $\Psi_{\mathrm{crit}}$ start oscillating.

Concerning the question of inner wind solutions two different types of solutions are possible. Either we insist on every solution to be a critical one regardless of whether it is global or not, or we argue that the crucial feature of a wind is to be global regardless of being supermagnetosonic. 

In the first case a flux surface can be found for which the acceleration is not sufficient for the plasma to reach the asymptotic regime (for $\Psi=0.42\,\Psi_{\mathrm{crit}}$, with the parameter set $a_{\mathrm{disk}}=2.5R_*$, $A=10$, $\psi=0.1364$). In this case, the wind still reaches the fast magnetosonic point but is then so much decelerated  that it closes very quickly afterwards (upper panel in Fig.~\ref{fig:4a}), where $u_p^{\mathrm{end}}=u_{FM}^{\mathrm{end}}$. 
This implies that the inner part of the asymptotic jet is empty despite having a constant mass distribution for each $\Psi$ at $x_{\mathrm{inj}}$. Even though the flux surface extends to some asymptotic radius, the plasma flow stops somewhere on its way along $\Psi$. Strictly speaking, the supermagnetosonic non-global solutions cannot be considered to be physical solutions.
This behaviour is due to the fact that we do not take pressure into consideration and have no self-consistent solutions. Especially close to the z-axis a wind can no longer be centrifugally driven and Lorentz-forces become negligible, hence pressure becomes more important. A mainly Poynting-flux driven cold wind (see Eq. (\ref{equ:46})), where $RB_{\phi} \rightarrow 0$ on the z-axis, will obviously not have any mechanism by which it can be accelerated away from the star. Only those winds can reach the asymptotic radius for which the flux tube functions $\Phi(R)$ decrease over a sufficiently long range. All other winds are not sufficiently accelerated. 

On the other hand, we can argue that because of the small influence of centrifugal and Lorentz-forces near the z-axis a wind becomes less accelerated so that it can no longer reach the magnetosonic point. This wind flows only with submagnetosonic velocities, but reaches the asymptotic jet radius. Hence we look for the smallest energy possible for which the wind can be global (Fig.~\ref{fig:4a}). 
\begin{figure}
\resizebox{\hsize}{!}{\includegraphics{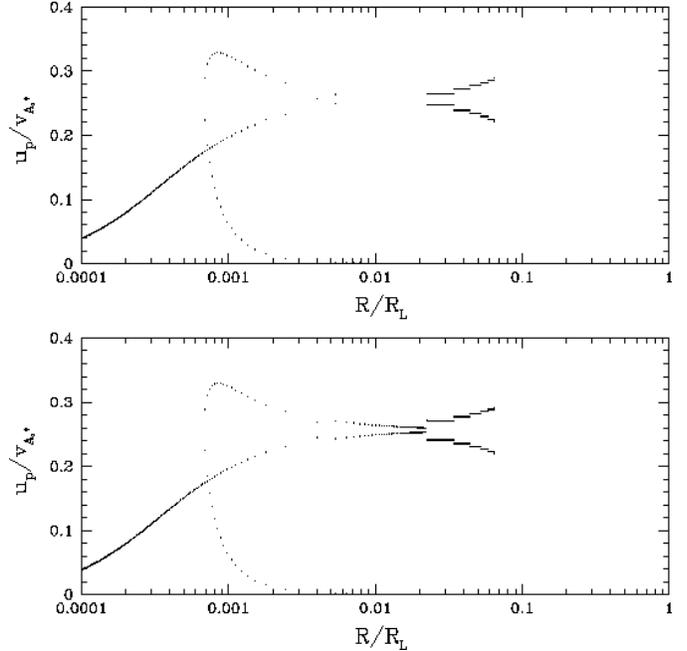}}
\caption{Poloidal wind velocity $u_{p}$. The solutions are along the flux surface $\Psi=0.42$ with $a_{\mathrm{disk}}=2.5R_*$, $\psi=0.1364$. The critical non-global solution for \mbox{$E_{\mathrm{crit}}=1.0000005203$} is shown in the upper panel, the submagnetosonic global one for $E_{\mathrm{crit}}=1.0000005215$ in the lower panel. Here the physical solution is the lower branch }
\label{fig:4a}
\end{figure}
In this case the physical wind solution is the lower branch. We claim that the energy continues to decrease smoothly from one flux surface to the next as is the case for the critical solutions. This feature is found for the supermagnetosonic, global solutions and should be maintained. For small $\sigma_*$ (\cite{fendt96}) the principal acceleration of the plasma occurs at radii smaller than the Alfv\'en-radius. There the plasma and the magnetic field corotate. This means that the acceleration is mainly due to the centrifugal force which decreases towards the z-axis. Consequently, less energy can be expected to be available near the z-axis to drive the wind.
The final conclusion is that we have to investigate a hot wind in the future for describing the physics more consistently.

\subsection{The influence of the disk radius $a_{\mathrm{disk}}$}\label{chap:61}
We investigate the features of the wind for a constant value of the unnormalized critical surface $\Psi_{\mathrm{crit}}^{\mathrm{unnorm}}$  (Examples I-III in Table \ref{tab:2}; realistic disk radii are at about $a_{\mathrm{disk}}\approx(2-3)R_{*}$). This means we keep $\psi/a_{\mathrm{disk}}=\mathrm{const}$, varying $\psi$ and $a_{\mathrm{disk}}$ at the same time, and implies that the amount of flux passing through the hole between the star and disk and the asymptotic jet radius $R_{\mathrm{jet}}=7937R_*$ remain unchanged as well. Figure \ref{fig:9} shows the influence for three different choices of $/psi-a_{\mathrm{disk}}$ on $\Psi$, $\Phi$ and $u_p$. For all three cases investigated, $\Phi$ has the same value at $x_{\mathrm{inj}}$. But the lower $\psi$ and $a_{\mathrm{disk}}$ are, the lower are the values of $\Phi$ (lower $\psi$ corresponding to less steep flux surfaces $\Psi$), and these are reached at successively smaller radii (for successively smaller $a_{\mathrm{disk}}$). For all three parameter sets, the solutions for $\Phi$ merge at a certain value of $x$, and thereafter track each other, remaining constant over the same length before they ultimately decrease to the same final value at $x=x_{\mathrm{jet}}$. This behaviour results in the wind reaching the fast magnetosonic velocity at smaller radii for lower values of $\psi$ and $a_{\mathrm{disk}}$. In addition, less energy is required to drive the wind, but with a smaller asymptotic velocity. We can conclude that a decrease of $\Phi$ over a long distance is more important for obtaining large asymptotic velocities than a larger difference between $\Phi(R_{\mathrm{inj}})$ and the minimum $\Phi_{\mathrm{min}}$ (Table \ref{tab:2}, Fig.~\ref{fig:9}). 
The step between obtaining a critical, global solution or only a non-global one takes place between $\Psi=0.42\Psi_{\mathrm{crit}}$ and $\Psi=0.49\Psi_{\mathrm{crit}}$. As already mentioned the wind solutions start to oscillate for flux surfaces which are lower than $50\%\Psi_{\mathrm{crit}}$. This is true for all three parameter sets. Hence the exact determination of where this transition takes place is not always possible.
As can be expected from an investigation of homogeneous solutions of the Grad-Shafranov equation, the parameter choices influence the physics most near the central source.

\begin{table}[h]  
\caption{{\small Three different parameter sets I, II and III for $\Psi_{\mathrm{crit}}^{\mathrm{unnorm}}=\mathrm{const}$}. $E_{\mathrm{crit}}-1$ refers to $\Psi_{\mathrm{crit}}=1.00$}
\label{tab:2}
\begin{center}
\begin{tabular}{|l|c|c|l|}
\hline
 &  & & \\[-.3cm]
 & I & II & III \\
\hline
 &  & & \\[-.3cm]
$\psi$ & 0.1364 & 0.2727 & 0.6000 \\
$a_{\mathrm{disk}}$ $(R_*)$ & 2.5 & 5.0 & 11.0 \\
$(E_{\mathrm{crit}}-1)$ $(10^{-5})$ & $1.232$ & $1.234$ & $1.413$\\
$u_{p}^{\infty}$(km s$^{-1}$) & 444 & 445 & 482\\
\hline
\end{tabular}
\end{center}
\end{table}

\begin{figure}
\resizebox{\hsize}{!}{\includegraphics{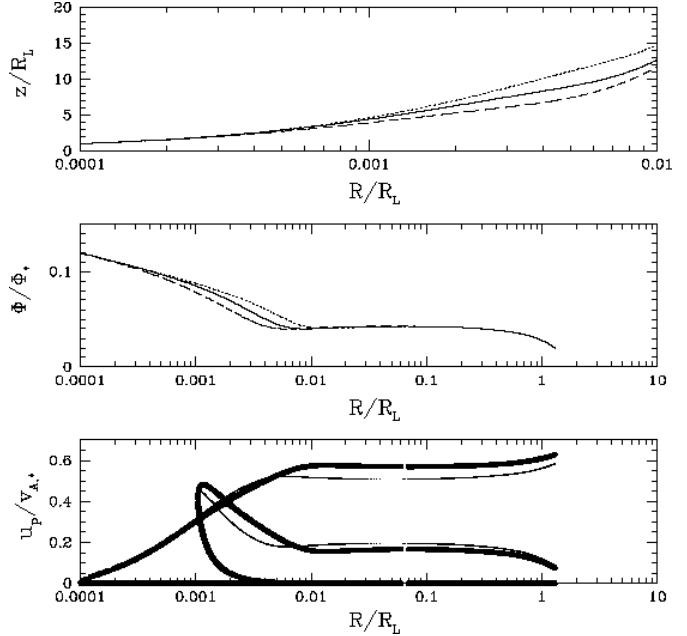}}
\caption{Flux surface (upper panel) $\Psi=0.86$ and flux tube function (middle panel) for $a_{\mathrm{disk}}=11$, $\psi=0.6$ (dotted curve),  $a_{\mathrm{disk}}=5$, $\psi=0.2727$ (thick curve) and for $a_{\mathrm{disk}}=2.5$, $\psi=0.1364$ (dashed curve). The lower panel shows the poloidal velocity for $a_{\mathrm{disk}}=11$, $\psi=0.6$ (thick curve) and $a_{\mathrm{disk}}=2.5$, $psi=0.1364$ (thin curve)}
\label{fig:9}
\end{figure}

\subsection{The weighting between $\Psi^{\mathrm{hom}}$ and $\Psi^{\mathrm{spec}}$}\label{sec:weight}\label{chap:63}
Now we investigate the influence of the constant $A$ of Eq. (\ref{equ:10}) on the wind solutions. All other parameters are kept fixed and all solutions are for $\Psi=\Psi_{\mathrm{crit}}$. Figure \ref{fig:7} shows the characteristics of the flux surfaces and the flux tube functions for $A=5, 10, 120$ and $240$. A larger A-value has the same effect as increasing the integration constant $\psi$: more flux can escape to infinity and less is forced to pass through the hole between the star and the disk. 
The position of the asymptotic radius is shifted to higher values for decreasing $A$. For example for $A=10$ the critical flux surface will collimate at radii about five times larger than for $A=120$. This is due to the larger impact of $\Psi^{\mathrm{hom}}$ on the total solution which will shift the asymptotic radius to a higher value.
The higher the value of $A$, the greater $\Phi(x_{\mathrm{inj}})$ is and its rate of decrease for small radii, and the closer to $a_{\mathrm{disk}}$ does the critical surface reach its minimum $\Phi$-value. This results in a wind reaching higher velocities near the star, and the fast magnetosonic velocity becomes larger for larger $A$-values (Fig.~\ref{fig:8a}). In contrary to the results in Sect. \ref{chap:61}, the maximum velocity is higher when the flux tube function decreases more rapidly but for a shorter distance (compare $A=240$ to $A=120$). Here the stronger decrease of $\Phi$ for $A=240$ overcomes the fact that it only decreases for a shorter distance. After $\Phi$ reaches its minimum value, the flux tube function increases for all $A$ resulting in a deceleration of the wind. But for intermediate radii the features of the flux tube function vary depending on $A$. For small $A$, $\Phi$ becomes constant and even decreases in the collimation region. For larger $A$ the flux tube function is increasing. Again there is a decrease in the collimation region. Nevertheless this continuing increase of $\Phi$ for intermediate radii is crucial for the wind to be able to reach the asymptotic radius.
For $A=120$ the increase in the flux tube function is not so steep that a global wind solution could no longer be found. On the other hand, for $A=240$ no global solution any longer exists despite the flux tube function having qualitatively the same appearance as that for $A=120$. 
\begin{figure}
\resizebox{\hsize}{!}{\includegraphics{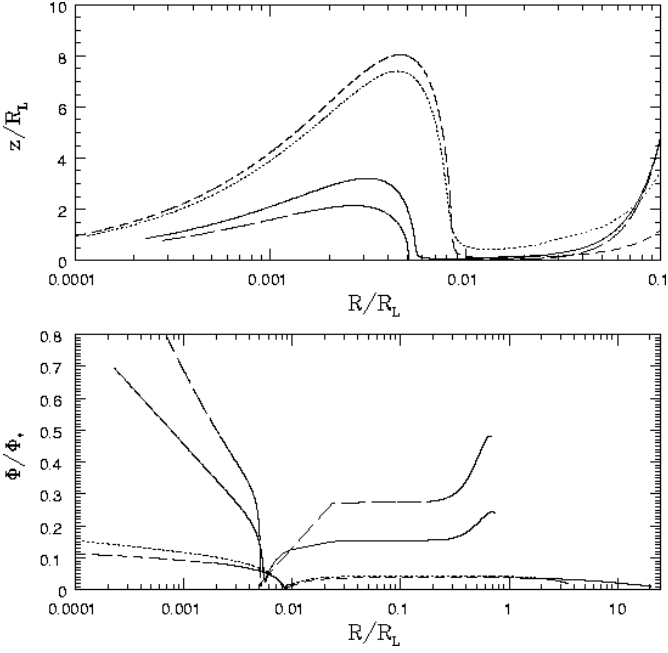}}
\caption{The flux surfaces (upper panel) and the flux tube function (lower panel) for $A=240$ (long--dashed), $A=120$ (solid), $A=10$ (dotted) and $A=5$ (short--dashed). All calculations are done for the critical flux surface, $\Psi=1.00$, $a_{\mathrm{disk}}=11R_*$ and $\psi=0.6$}
\label{fig:7}
\end{figure}
\begin{figure}
\resizebox{\hsize}{!}{\includegraphics{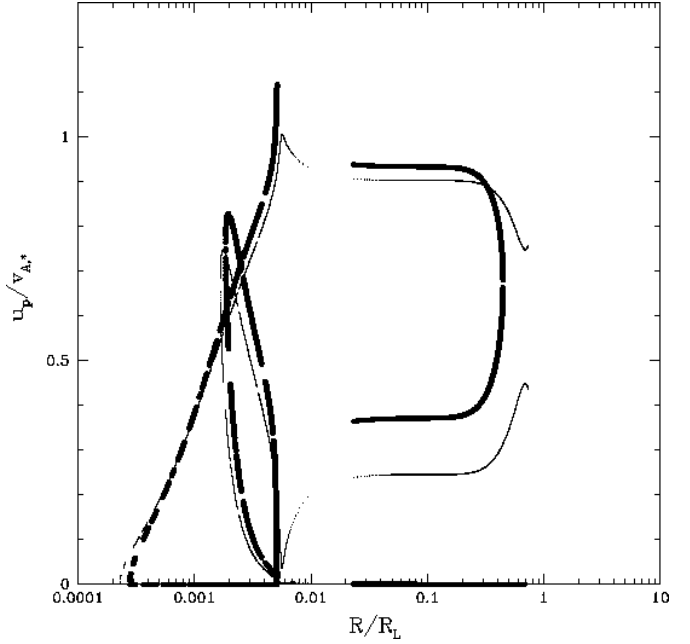}}
\caption{The thin curve shows the poloidal wind for $A=120$, the thick curve that for $A=240$. It is  $\Psi_{\mathrm{crit}}$, $a_{\mathrm{disk}}=11R_*$ and $\psi=0.6$}
\label{fig:8a}
\end{figure}

\subsection{The role of the flux integration constant $\psi$}\label{chap:64}
The integration constant $\psi$ represents the ability of the homogeneous solution for the flux surfaces to open and to let flux escape to infinity. The higher $\psi$ is, the higher the bending of the dipole-like shaped field line after its opening near $a_{\mathrm{disk}}$. For the infinitely small disk this occurs nearly at $z=0$.
For $\psi=0$ Eq. (\ref{equ:7}) will not contribute to radii $R>a_{\mathrm{disk}}$ and the magnetosphere is forced to go through the hole between star and disk. We get a dipole which is squeezed by the inner edge of the disk. But due to the special solution (Eq. (\ref{equ:10})) plasma can escape to larger radii. Then the shape of the outer flux tube function is only determined by $\Psi^{\mathrm{spec}}$, consequently $R_{\mathrm{jet}}\equiv a$. For asymptotic radii the flux tube function increases up to the final jet radius, which is in contrast to the solutions we get for $\psi=0.6$ (see the result for $A=10$ in Sect. \ref{chap:63}), where the flux tube functions decrease for asymptotic radii. For both cases the wind can reach the asymptotic region.

The influence of $\psi$ on the shape of the magnetic flux surfaces is discussed now for five different values, as listed in Table \ref{tab:3} and plotted in Fig.~\ref{fig:10},  and for the case of the critical surface $\Psi_{\mathrm{crit}}$. For $\psi=2.0$ the flux surface opens at $R_0\approx 4.5R_*=a_{\mathrm{disk}}$ and then keeps very close and flat along the surface of the disk. For $\psi=2.5$ the flux surface dips down to $z=0$ at $R_0\approx 4.0R_*<a_{\mathrm{disk}}$, then rises instantly, but only to a height much smaller than that which the dipole reaches as a maximum, and becomes constant about $1R_*$ later. Increasing $\psi$ to 3.5, $R_0$ is further lowered to about  $3.5R_*$. Afterwards the flux surfaces rise very quickly without having a constant part parallel to the x-axis. These characteristics occur and are enhanced for  $\psi=4.5$ and $\psi=10$. For these examples the normalisation factor $A=10$ is used. For smaller values of $A$ the influence of the homogeneous solution $\Psi^{\mathrm{hom}}$ is so great that the flux surfaces would collimate at much smaller radii (e.g. for $A=10$ they collimate at about 8000 $R_*$, for $A=120$ at about 1650 $R_*$).
\begin{table}
\caption{Five different values for the integration constant C and the related radius $R_0$, at which the critical flux surface opens}
\label{tab:3}
\begin{center}
\begin{tabular}{|l|c|c|c|c|l|}
\hline
 &  & & \\[-.3cm]
& I & II & III & IV &  V\\
\hline
 &  & & \\[-.3cm]
$\psi$ & 2.0 & 2.5 & 3.5 & 4.5 & 10 \\
$\Psi_{\mathrm{crit}}^{\mathrm{unnorm}}$ & 1.32  & 1.37 & 1.43 & 1.48 & 1.64 \\
$R_0$ ($R_*$) $\approx$ & 4.5 & 4.0 & 3.5 & 3.0 & $<$ 3.0\\
\hline
\end{tabular}
\end{center}
\end{table}

\begin{figure}
 \resizebox{\hsize}{!}{\includegraphics{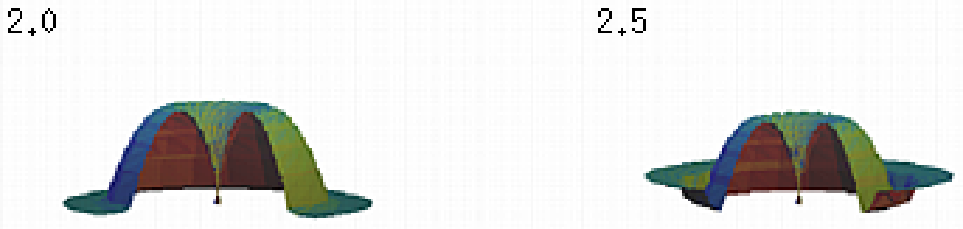}}
 \resizebox{\hsize}{!}{\includegraphics{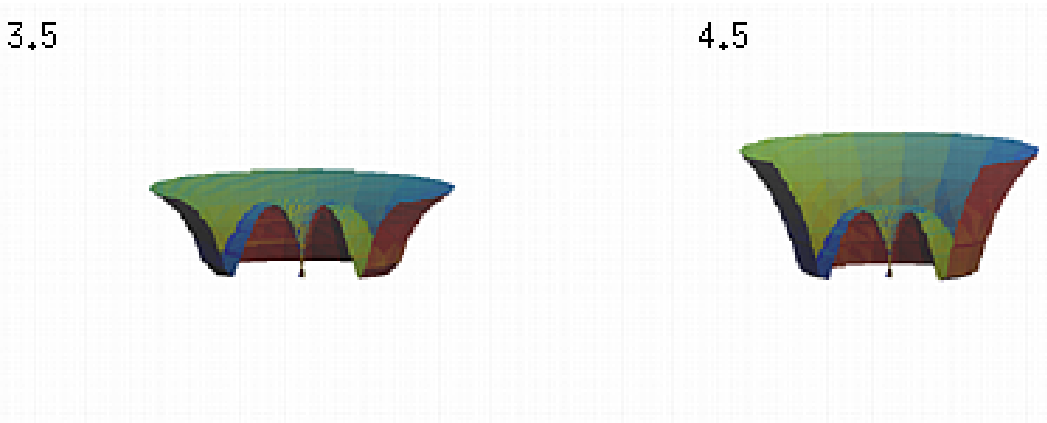}}
 \resizebox{\hsize}{!}{\includegraphics{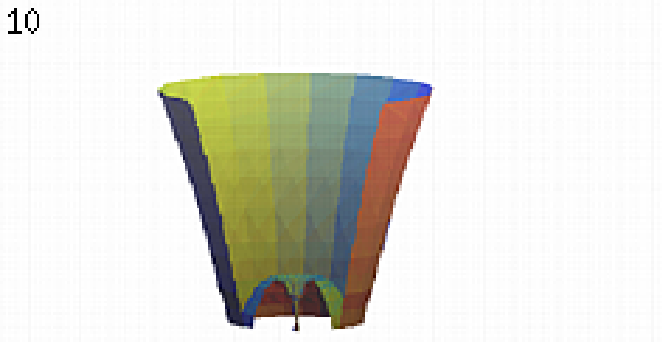}}
\caption{The critical flux surfaces for $\psi=2.0$, 2.5, 3.5, 4.5 and 10 and  $a_{\mathrm{disk}}=4.5R_*$, $A=10$. All plots have the same scaling and are shown from $R=(0-6)R_*$}
\label{fig:10}
\end{figure}
\subsection{The role of gravity}
We next consider the role of gravity on the wind. Note that in principle the vacuum solution $\Psi^{\mathrm{hom}}$ should now be derived from the Grad-Shafranov equation (Eq. (\ref{equ:81})) including the gravitational redshift $\alpha$. But it can be shown that the influence of $\alpha$ for the solution is negligible (see Appendix C).

The importance of the redshift factor $\alpha$ on the wind is seen at small radii. A wind can only emanate from the star if $(E-1) > 0$. 
From Eq. (\ref{equ:ener}) we recognize that for $\alpha<1$ solutions with positive energy exist only for $u_p(x_{\mathrm{inj}})>0$ which cannot be satisfied by a cold wind ($p_{*}\ll \epsilon$). This also means that the critical energy can no longer be calculated by requiring $m_0(x_{\mathrm{inj}})=0$ (see Eq. (\ref{equ:79})) because this implies $u_p(x_{\mathrm{inj}})=0$. We can conclude that the cold limit is not a good approximation for our problem. The deviation from $\alpha=1$ by taking gravitation into consideration is of the order $10^{-6}$. That is the same order of magnitude as for the expression $\epsilon/(1-\epsilon)$.
\begin{figure}
\resizebox{\hsize}{!}{\includegraphics{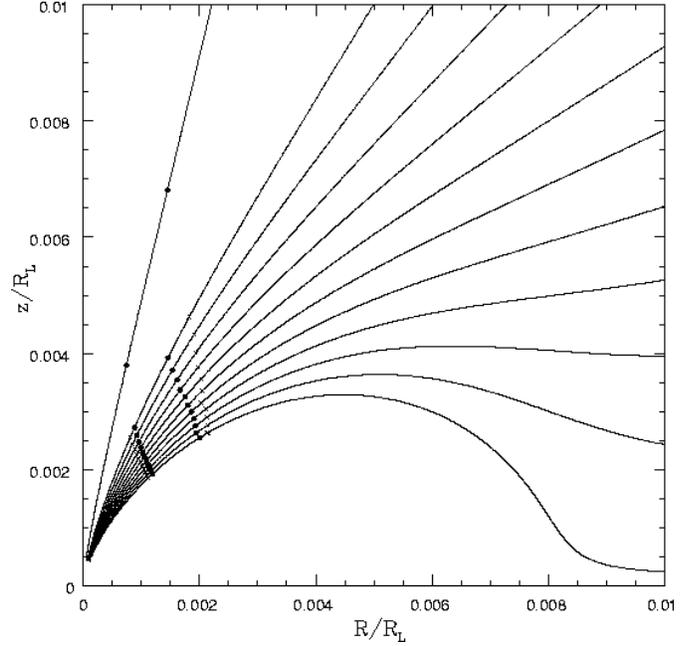}}
\caption{Magnetic flux surfaces for $a_{\mathrm{disk}}=11R_*$ and $\psi=0.6$. The inner dots represent the positions of the Alfv\'en points and the outer ones those of the fast magnetosonic points, for the wind without gravity. The inner crosses show the positions of the Alfv\'en points and the outer ones those of the fast magnetosonic points, for the wind with gravity. The contour levels are $\Psi=1.0$ to $0.64$ in steps of $0.036$ and $\Psi=0.49$, 0.57}
\label{fig:4}
\end{figure}

\begin{figure}
\resizebox{\hsize}{!}{\includegraphics{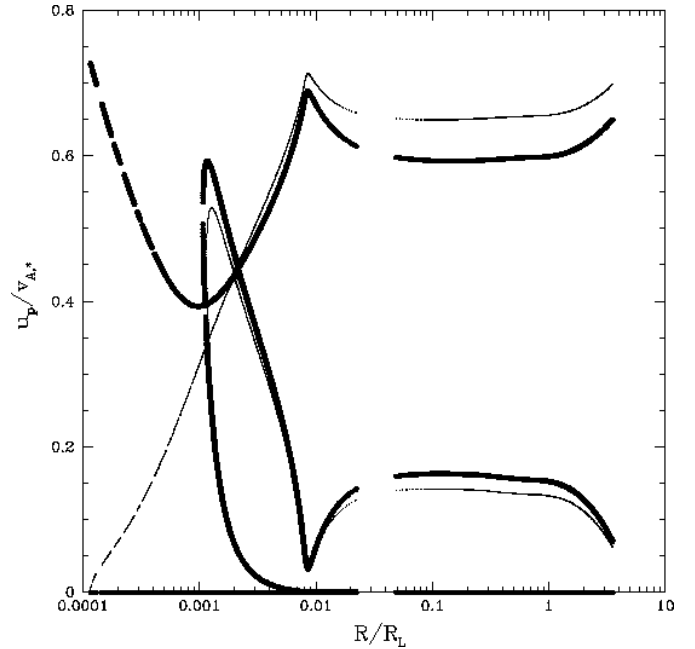}}
\caption{Poloidal wind velocity $u_{p}$. The solution is along the flux surface $\Psi=1$. The thin curve is calculated with $\alpha=1$, the thick one with $\alpha<1$. The inner disk radius is $a_{\mathrm{disk}}=11R_*$ and $\psi=0.6$}
\label{fig:5}
\end{figure}
Figures \ref{fig:4} and \ref{fig:5} demonstrate the main differences between a wind calculated for the non-gravity limit ($\alpha=1$) and the case including gravity. Very close to the star ($\alpha<1$) the non-Newtonian wind cannot start with zero poloidal velocity. We claim that the ratio of sound speed to Alfv\'en velocity on the stellar surface is $c_{s,*}/v_{A,*}=u_p(x_{\mathrm{inj}})/v_{A,*}<1$. A sound speed of $c_{s,*}\simeq100$ km s$^{-1}$ in a stellar corona can be assumed. As can be seen in Fig.~\ref{fig:5} this ratio is fairly higher in our calculation (about 0.73 for $\epsilon=1.26\,10^{-6}$). Instead of determining the critical energy and $\epsilon$ for a given injection radius $x_{\mathrm{inj}}$, we can choose a higher value for $\epsilon$ and still obtain a critical solution starting at about the same injection radius but with a smaller velocity. Along $\Psi=1$ reasonable results can be obtained up to $\epsilon=1.8\times10^{-6}$ and $c_{s,*}/v_{A,*}=0.55$. This value is still too high.

The wind reaches the asymptotic radius but for lower velocities due to the deceleration caused by gravity. As a consequence, the value of the critical energy is smaller. 
In comparison with the case of a constant $\alpha$, the Alfv\'en-point for a radial dependent $\alpha(r)$ for all flux tube functions $\Psi$  is reached at smaller radii, and the fast magnetosonic point at larger radii. Their shapes are nonetheless qualitatively similar. For a calculation with or without gravitation, the  Alfv\'en-points as well as the fast magnetosonic points move linearly  inward with decreasing $\Psi$.
 Figure \ref{fig:4} also shows that for $\alpha<1$ a critical wind solution cannot be found for flux tube functions as small as for  $\alpha=1$, i.e. the Alfv\'en-point and the fast magnetosonic point can no longer be determined. 
The position of the Alfv\'en- surface and the fast magnetosonic surface stop rising linearly on the surfaces for which $\Psi\simeq 0.57\Psi_{\mathrm{crit}}$. They appear to run parallel to the flux surfaces, i.e. the z--axis, a result also obtained by Fendt \& Camenzind (1996).
For both calculations the fast magnetosonic points are at radii smaller than the disk radius.  
For the parameter set (I) of Table \ref{tab:2} a wind calculated along the flux tube function $\Psi=0.57$ and including the gravitation of the star can no longer reach the asymptotic region. 

\subsection{Derived wind parameters}
Figure \ref{fig:14} shows the poloidal velocity for $\Psi=0.89$ and the corresponding Alfv\'en-Mach number $M$, the poloidal current $I$, the toroidal velocity $u_{\phi}$ and the density $n$. A constant poloidal current $I$ is equivalent to a force-free magnetosphere. But the current strength decreases towards larger radii by a factor of about $7-8$.
The relation between $u_{p}$ and $u_{\phi}$ is a subject of interest when the observed emission lines are to be interpreted. There is a high  velocity dispersion if $u_{p}$ and $u_{\phi}$ are of about the same order of magnitude, hence the spectral lines are strongly Doppler broadened. In our model these two velocities are of about the same order of magnitude for small radii. At the Alfv\'en point, one finds $u_p/u_{\phi}\sim 0.7$, at the fast magnetosonic point $u_p/u_{\phi}\sim 3.7$ and at the asymptotic radius $u_p/u_{\phi}\sim 2.2\times 10^{4}$.  
The plasma density between the stellar position and the asymptotic jet radius decreases by several orders of magnitude.

\begin{figure*}
\begin{center}
\resizebox{\hsize}{!}{\includegraphics{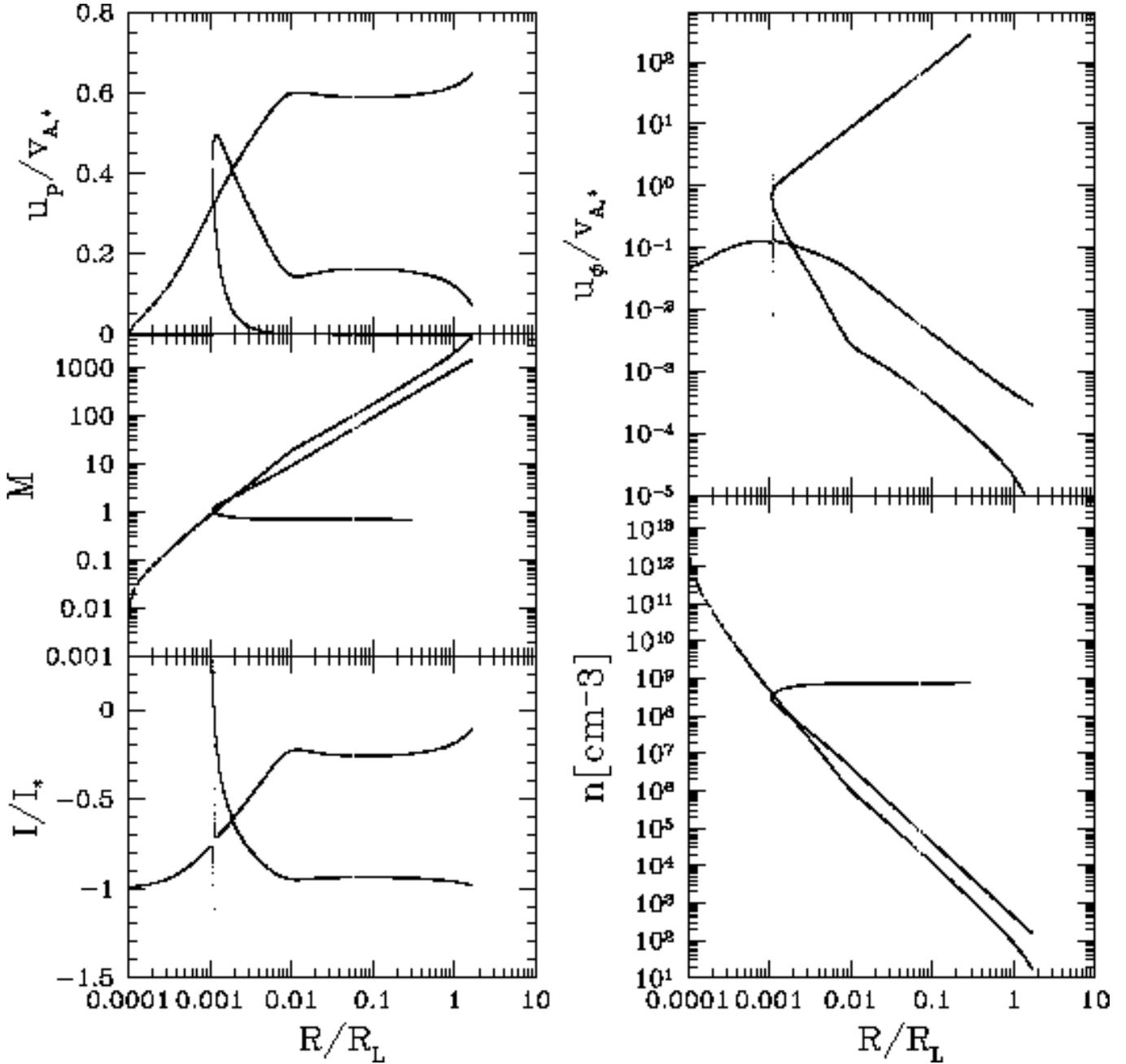}}
\end{center}
\caption{Poloidal wind velocity $u_{p}$, Alfv\'en Mach number $M$, current $I$, toroidal wind velocity $u_{\phi}$ and density $n$. The solution is along the flux surface $\Psi=0.89$. All curves are calculated for $\alpha=1$, $a_{\mathrm{disk}}=11R_*$ and $\psi=0.6$}
\label{fig:14}
\end{figure*}

\section{Conclusions}\label{chap:7}
In this work we presented an analytical model for the magnetic flux surfaces which reproduces the main features of a protostellar wind that collimates at large radii. The complete expression for the flux surface consists of three different parts which determine the solution for small, intermediate and large radii. For the region near the star a stellar dipole field is assumed and calculated by extending the computation of Kundt \& Robnik (1980). We assume the existence of an infinitely thin disk perpendicular to the rotation axis at a distance of a few stellar radii ($a_{\mathrm{disk}}=11R_*$ for most examples). The stellar dipole field turns out to be essential for the acceleration of the wind to supermagnetosonic velocities, which occurs very near the star. For intermediate radii the magnetosphere is nearly spherical and may be described as monopole. Collimation of the field lines is achieved at radii $R_{\mathrm{jet}}\approx (3-4)R_{L}$. This region is characterized by magnetospheres similar to those for force-free solutions (Appl \& Camenzind 1993b).  

We investigate the dynamics of a wind in these collimated magnetospheres for
 axisymmetric and stationary flows, obtaining a 2D picture of the wind. As a first approximation we consider a 
cold wind, neglecting the plasma pressure. The physical solution must start at zero poloidal velocity and pass smoothly through the Alfv\'en point and
 the fast magnetosonic point. 
The resulting wind solutions reach the jet radius in the Newtonian limit. The asymptotic jet velocities of about a few hundred km/s are comparable to those observed and depend mainly on the choice of the magnetization parameter $\sigma_*$.
The asymptotic velocity of the outflowing plasma decreases the closer it is to the jet axis. Realistic asymptotic jet velocities are obtained for the parameters $\sigma_*=10^{-8}$, $a=1430R_*, A=10, \psi=0.6, a_{\mathrm{disk}}=11R_*$ and $u_p^\infty=482$ km s$^{-1}$ for $\Psi_{\mathrm{crit}}$. The influence of the various parameters of the model on the wind solutions is discussed in more detail in the text.
If we include the effects of gravity by means of the redshift factor which accounts for the gravitational force of the central star, solutions must start with a non-vanishing velocity. This is equivalent to a rise in the sound speed up to about $c_{S,*}/v_{A,*} < 1$. Such a high ratio of sound to stellar Alfv\'en velocity could only be explained by temperatures near $10^7$ K. By contrast, shock heating due to plasma accretion onto the star allows only temperatures up to $10^6$ K. Hence an additional source of heating is required. Another explanation might be that a hot wind is essential to obtain a solution.  

For plasma flowing along magnetic flux surfaces which are close to the z-axis, critical wind solutions can no longer be determined. Either there is no global solution branch extending to the asymptotic radius of the flux surface, or there is a global plasma flow which remains submagnetosonic throughout the entire flow.
The last two points clearly demonstrate the need to calculate a hot wind instead.

The velocity field throughout the jet and the related density and temperature obtained here can be used for future work as input to calculate forbidden emission line profiles of T Tauri Stars (\cite{solf93}). These forbidden emission lines are a powerful means of probing the outflow of Classical T Tauri Stars and of studying the link between the ejection and accretion processes (Breitmoser \& Camenzind 2000).
 
\begin{acknowledgements}
This work is supported by the Deutsche Forschungsgemeinschaft through the Priority Research Program "Physics of Star Formation". EB thanks Dr.~A.~Meiksin for comments improving the clarity of the presentation.
\end{acknowledgements}

\begin{appendix}

\section{Homogeneous solution of the Grad-Shafranov equation}\label{model} 

A homogeneous solution of the Grad-Shafranov equation, written in cylindrical coordinates,
\begin{equation}
{\partial^2 A_{\varphi}\over{\partial R^2}}+{\partial^2 A_{\varphi}\over{\partial z^2}}+\frac{1}{R}
{\partial A_{\varphi}\over{\partial R}}-\frac{A_{\varphi}}{R^2}=0,\label{equ:b1}
\end{equation}
can be obtained by separation of variables
\begin{equation}
A_{\varphi}(R,z) = \int_0^\infty b(k)\,J_1(kR)\,\exp(-k\vert z\vert)\,dk
\end{equation}
given by suitable boundary conditions. The magnetic flux surfaces are given by $\Psi=RA_{\varphi}$, and $J_1(kR)$ is a Bessel function of first kind and first order. We make the following assumptions:
\begin{itemize}
\item  The plasma is perfectly conducting, consequently no field lines can penetrate the disk and the disk is screened by induced surface currents.
\item The disk is infinitely small and exists therefore only at $\mathrm{z}=0$.
\item The dipole is perpendicular to the disk.
\item The problem is axisymmetric hence the induced vector potential $A$ is given by $A_{\varphi}$ alone (\cite{kundt}). 
\end{itemize}
The two boundary conditions are 

\begin{itemize}
\item $B_{R}(R,z=0) = 0$ for $R_{*} \le R \le a_{\mathrm{disk}}$ 
\item $B_{z}(R,z=0) = 0$ for $R > a_{\mathrm{disk}}$.

\end{itemize}

For the most general case of open flux surfaces the second condition, in contrast to Kundt \& Robnik (1980),  reads
\begin{eqnarray}
-A_{\varphi}^{\mathrm{Dipole}}(R,0)+\frac{\mathrm{Const}}{R}&=&\int_{0}^{\infty}b(k)\,J_1(kR)\,dk, \nonumber \\
& &  \hbox{for} \quad R > a_{\mathrm{disk}}. 
\end{eqnarray}
In addition to their result we get a part proportional to the integration constant $\mathrm{Const}$
\begin{eqnarray}
b(k) &=& -\sqrt{\frac{2k}{\pi}}\int_{a_{\mathrm{disk}}}^{\infty}t^{3/2}J_{3/2}(kt)\,dt
 \frac{d}{dt}\int_{t}^{\infty}\frac{\mathrm{Const}}{\tau\sqrt{\tau^2-t^2}}d\tau \nonumber \\
 &=&\mathrm{Const}\int_{a_{\mathrm{disk}}}^{\infty}\left(\frac{\sin(kt)}{kt^2}-\frac{\cos(kt)}{t}\right)dt
\end{eqnarray}
and
\begin{eqnarray}
A_{\varphi}&=&\left. \frac{\mathrm{Const}}{R}\mathrm{Im}\frac{-\sqrt{-t^2-2i|z|t+R^2}}{t}\right]_{t=a_{\mathrm{disk}}}^{\infty} \nonumber \\
&=&\frac{\mathrm{Const}}{\sin\theta}\left(\frac{1}{\rho}-\gamma\right).
\end{eqnarray} 
For our further calculation we use $\psi=\mathrm{Const}\, a_{\mathrm{disk}}\pi/2\mu_*$. 

\section{Magnetic structure of collimated magnetospheres}
\label{app:1}
For the magnetic fields $B=B^{\mathrm{hom}}+B^{\mathrm{spec}}$ we get 
\begin{eqnarray}
B_r &=& {1\over{r^2\sin\theta}}\,{{\partial\Psi}\over{\partial\theta}}, \\
B_\theta &=& -{1\over{r\sin\theta}}\,{{\partial\Psi}\over{\partial r}},\\
B_{r}^{\mathrm{hom}}&=&\frac{4 \mu_*\, \cos\theta}{\pi\, r^3}\left(\arctan(\gamma)+\frac{1}{\gamma}\left(1-\delta+C\frac{\delta}{2}\right)\right),\nonumber\\
& & \\
B_{\theta}^{\mathrm{hom}}&=&\frac{2\mu_* \, \sin\theta}{\pi\, r^3}\left(\arctan(\gamma)-\frac{\cot^2\theta}{\gamma}+\frac{\gamma\delta(1+\rho^2)}{\sin^2\theta}\right. \nonumber\\
& & +C\left. \frac{\rho^2\gamma}{\sin^2\theta}(1-\delta(1+\gamma^2))\right),\\
B_{r}^{\mathrm{spec}}&=&A\,\frac{a^2+2\, r^2 \cos\theta}{r^2 R_{c}^2}\frac{1}{\mathrm{arg}}, \\
B_{\theta}^{\mathrm{spec}}&=&-A\,\frac{2\, \sin\theta}{R_{c}^2}\frac{1}{\mathrm{arg}},
\end{eqnarray}
where
\begin{equation}
\delta:=(1+2\gamma^2-\rho^{-2})^{-1},
\end{equation}
\begin{equation}
\mathrm{arg}:=1+\frac{a^2}{R_{c}^2}(1-\cos\theta)+\frac{r^2\sin^2\theta}{R_{c}^2}
\end{equation}
and $\nabla\cdot\vec{B}=0$.

\section{The Grad-Shafranov equation including the gravitational redshift}\label{app:3} 
In comparison to Eq. (\ref{equ:b1}) the Grad-Shafranov equation including the gravitational redshift (Eq. (\ref{equ:81})) reads
\begin{equation}
{\partial^2 A_{\varphi}\over{\partial R^2}}+{\partial^2 A_{\varphi}\over{\partial z^2}}+(\frac{1}{R}+\frac{B}{R^2})
{\partial A_{\varphi}\over{\partial R}}+(-\frac{1}{R^2}+\frac{B}{R^3})A_{\varphi}=0,\label{equ:c1}
\end{equation}
with
\beq
B=\frac{GM_*}{r\alpha c^2}.
\eeq
Eq. (\ref{equ:c1}) differs from Eq. (\ref{equ:b1}) in the dimensionless
expression $B/R$. This has a maximum value at $R=R_*$. An estimate shows that
 $B/R_*\approx 10^{-11}$ is negligible for the calculation of $\Psi$.
\end{appendix}

\begin{thebibliography}{cam97}{}
\bibitem[Aly 1980]{aly} Aly, J.J., 1980, A\&A 86, 192
\bibitem{appl93} Appl, S., Camenzind, M., 1993a, A\&A 270, 71
\bibitem{appl93b} Appl, S., Camenzind, M., 1993b, A\&A 274, 699
\bibitem{beskin} Beskin, V., Pariev, V.I., 1993, Phys. Uspekhi 36, 529
\bibitem{bland} Blandford, R.D., Znajek, R.L., 1976 MNRAS 179, 433
\bibitem{ebII}Breitmoser, E., Camenzind, M. 2000, in preparation
\bibitem{cam86a} Camenzind, M., 1986a, A\&A 156, 137 
\bibitem{cam86b} Camenzind, M., 1986b, A\&A 162, 32
\bibitem{cam87} Camenzind, M., 1987, A\&A 184, 341
\bibitem[Camenzind 1990]{cam90} Camenzind, M., 1990, Rev. Modern Astron., 3, 234
\bibitem{cam96} Camenzind, M., 1996, in {\it Solar and Astrophysical
Magnetohydrodynamic Flows}, ed. K.C. Tsinganos, Kluwer (Dordrecht), p. 699
\bibitem{cam97} Camenzind, M. 1997, in IAU Symp. 182 {\it Herbig--Haro
Flows and the Birth of Low Mass Stars}, eds. B. Reipurth and C. Bertout,
Kluwer (Dordrecht), p. 241
\bibitem[Fendt et al. 1995]{fendt95} Fendt, C., Camenzind, M. Appl, S., 1995, A\&A 300, 791
\bibitem[Fendt \& Camenzind 1996]{fendt96} Fendt, C., Camenzind, M., 1996, A\&A 313, 591
\bibitem{fendt97} Fendt, C., 1997, A\&A 319, 1025
\bibitem{Freidberg} Freidberg, 1987, Ideal MHD, Plenum Press (N.Y.)
\bibitem{hein78} Heinemann, M., Olbert, S. 1978, J. Geophys. Res. 83, 2457
\bibitem[Heyvaerts \& Norman 1996]{heyvaerts96}Heyvaerts, J., Norman, C.A., 1996, in {\it Solar and Astrophysical Magnetohydrodynamic Flows}, ed. K.C. Tsinganos, Kluwer (Dordrecht), p. 459 
\bibitem{greg96} MacGregor, K.B., 1996, in {\it Solar and Astrophysical
Magnetohydrodynamic Flows}, ed. K.C. Tsinganos, Kluwer (Dordrecht), p. 301
\bibitem[Khanna 1998]{khanna98}Khanna, R., 1998, MNRAS 294, 673
\bibitem[Kundt \& Robnik 1980]{kundt} Kundt, W., Robnik, M., 1980, A\&A 91, 305
\bibitem[Lery et al. (1998)]{lery98}Lery, T., Heyvaerts, J., Appl, S., Norman, C.A., 1998, A\&A 337, 603
\bibitem[Lery et al. (1999)]{lery99}Lery, T., Heyvaerts, J., Appl, S., Norman, C.A., 1999, A\&A 347, 1055
\bibitem{mobarry86} Mobarry, C.M., Lovelace, R.V.E., 1986, ApJ 309, 455
\bibitem{okamoto} Okamoto, I., 1992, MNRAS 254, 192
\bibitem[Paatz 1994]{paatz94} Paatz, G. 1994, Ph.D. thesis, Univ. Heidelberg
\bibitem[Paatz \& Camenzind 1996]{paatz96} Paatz, G., Camenzind, M., 1996, A\&A 308, 77
\bibitem{pelletier} Pelletier, G., Pudritz, R.E., 1992, ApJ 394, 117
\bibitem[Ray et al. 1996]{Ray} Ray et al., 1996, ApJL 468, L103
\bibitem[Riffert 1980]{riffert} Riffert, H., 1980, Astrophys. Space Science 71, 195
\bibitem{rosso} Rosso, F., Pelletier, G., 1994, A\&A 287, 325
\bibitem{Sauty} Sauty, C., Tsinganos, K., 1994, A\&A 287, 893
\bibitem[Solf \& B\"ohm 1993]{solf93} Solf, J., B\"ohm, K.-H., 1993, ApJ Letters, 410, L31
\bibitem[Thorne et al. 1986]{thorne} Thorne, K.S., Price, R.H., McDonald, D.A., Suen, W.--M., Zhang, X.--H., 1986, in : {\it Black Holes: The Membrane Paradigm}, eds. Thorne, K.S., Price, R.H., McDonald, D.A., Yale Univ. Press, New Haven
\bibitem[Trussoni et al. 1996]{trussoni} Trussoni, E., Sauty, C., Tsinganos, K., 1996, in 
{\it Solar and Astrophysical Magnetohydrodynamic Flows}, ed. K.C. Tsinganos, Kluwer (Dordrecht), p. 383
\bibitem[Tsinganos et al. 1996]{tsin96}Tsinganos, K., Sauty, C., Surlantzis, G., Trussoni, E., Contopoulos, J. 1996, in {\it Solar and Astrophysical Magnetohydrodynamic Flows}, ed. K.C. Tsinganos, Kluwer (Dordrecht), p. 427
\bibitem{weber} Weber, E.J., Davis, L., 1967, ApJ 148, 217
\end{thebibliography}
\end{document}